\documentclass[journal]{IEEEtran}

\ifCLASSINFOpdf
\else
   \usepackage[dvips]{graphicx}
\fi
\usepackage{url}

\usepackage{graphicx}

\usepackage{makecell}
\usepackage{colortbl}
\usepackage{mathtools}
\usepackage{booktabs}
\usepackage{commath}
\usepackage{multirow}
\usepackage{amssymb}
\usepackage{siunitx}
\usepackage{amsthm}
\usepackage[ruled,linesnumbered]{algorithm2e}
\usepackage{epstopdf}

\usepackage{interval}
\usepackage{tikz}
\usepackage{circledsteps}
\usepackage[inline]{enumitem}
\usepackage{siunitx}
\usepackage{stackengine}
\usepackage{xcolor}
\usepackage[normalem]{ulem}
\newcommand\redsout{\bgroup\markoverwith{\textcolor{red}{\rule[0.5ex]{2pt}{1pt}}}\ULon}

\newcommand{\cb}{\cellcolor{blue!15}}
\usepackage[caption=false,font=scriptsize]{subfig}
\usepackage[export]{adjustbox}
 \usepackage{draftfigure}

\usepackage{xspace}
\makeatletter
\DeclareRobustCommand\onedot{\futurelet\@let@token\@onedot}
\def\@onedot{\ifx\@let@token.\else.\null\fi\xspace}
\def\eg{\emph{e.g}\onedot} 

\def\ie{\emph{i.e}\onedot}

\def\etal{\emph{et al}\onedot}
\makeatother
\usepackage[capitalize]{cleveref}
\crefname{section}{Sec.}{Secs.}
\Crefname{section}{Section}{Sections}
\Crefname{table}{Table}{Tables}
\crefname{table}{Tab.}{Tabs.}
\AtBeginEnvironment{appendices}{\crefalias{section}{appendix}}
\DeclareSIUnit\eps{EPS}
\DeclareSIUnit\fps{FPS}

\usepackage{xcolor}
\usepackage{pifont}
\newcommand{\cmark}{\cellcolor{red!15}\ding{51}}%
\newcommand{\xmark}{\cellcolor{blue!15}\ding{55}}%

\usepackage{threeparttable}
\usepackage{tikz}

\usepackage{fancyhdr}
\makeatletter
\let\ps@IEEEtitlepagestyle\ps@fancy
\makeatother
\begin{document}

\title{Neuromorphic Imaging with Super-Resolution}

\author{Pei Zhang,~\IEEEmembership{Member, IEEE}, Shuo Zhu,~\IEEEmembership{Member, IEEE}, Chutian Wang, \par Yaping Zhao,~\IEEEmembership{Student Member, IEEE}, and Edmund Y. Lam,~\IEEEmembership{Fellow, IEEE}%
\thanks{This work was supported in part by the Research Grants Council of Hong Kong SAR (GRF 17201620, 17200321) and by ACCESS --- AI Chip Center for Emerging Smart Systems, sponsored by InnoHK funding, Hong Kong SAR.}%
\thanks{The authors are with the Department of Electrical and Electronic Engineering, The University of Hong Kong, Pokfulam, Hong Kong SAR, China (e-mail: zhangpei@eee.hku.hk, zhushuo@hku.hk, ctwang@eee.hku.hk, zhaoyp@eee.hku.hk, elam@eee.hku.hk). Edmund Y. Lam is also affiliated with ACCESS --- AI Chip Center for Emerging Smart Systems, Hong Kong Science Park, Hong Kong SAR, China.}%
\thanks{Corresponding author: Edmund Y. Lam.}}

\maketitle

\begin{abstract}
Neuromorphic imaging is an emerging technique that imitates the human retina to sense variations in dynamic scenes. It responds to pixel-level brightness changes by asynchronous streaming events and boasts microsecond temporal precision over a high dynamic range, yielding blur-free recordings under extreme illumination. Nevertheless, this modality falls short in spatial resolution and leads to a low level of visual richness and clarity. Pursuing hardware upgrades is expensive and might cause compromised performance due to more burdens on computational requirements. Another option is to harness offline, plug-in-play super-resolution solutions. However, existing ones, which demand substantial sample volumes for lengthy training on massive computing resources, are largely restricted by real data availability owing to the current imperfect high-resolution devices, as well as the randomness and variability of motion. To tackle these challenges, we introduce the first self-supervised neuromorphic super-resolution prototype. It can be self-adaptive to per input source from any low-resolution camera to estimate an optimal, high-resolution counterpart of any scale, without the need of side knowledge and prior training. Evaluated on downstream tasks, such a simple yet effective method can obtain competitive results against the state-of-the-arts, significantly promoting flexibility but not sacrificing accuracy. It also delivers enhancements for inferior natural images and optical micrographs acquired under non-ideal imaging conditions, breaking through the limitations that are challenging to overcome with frame-based techniques. In the current landscape where the use of high-resolution cameras for event-based sensing remains an open debate, our solution is a cost-efficient and practical alternative, paving the way for more intelligent imaging systems.
\end{abstract}


\IEEEpeerreviewmaketitle

\section{Introduction}
\IEEEPARstart{N}{euromorphic} imaging mimics the neural architecture of the human retina to sense scene variations. It generates asynchronous, temporal streaming events in response to per-pixel brightness changes within a field of view, encoding visual information with a fast speed ($\sim\SI{10}{\us}$) over a high dynamic range ($\sim\SI{120}{\decibel}$)~\cite{gehrig2024nature,zhu2024pw}. This novel modality, which enjoys blur-free and low-power recordings of ultra-fast moving targets under extreme illumination, has made groundbreaking advancements across multiple fields, such as computational imaging~\cite{mangalwedhekar2023nn,zhu2024ol}, and machine vision~\cite{ramesh2021tcsvt,liu2022tcsvt,zhao2024icip}.

Despite remarkable features, existing neuromorphic cameras present insufficient spatial resolution and fall short in offering an equivalent degree of visual clarity as most frame cameras. Augmenting spatial resolution at the hardware level is cost-prohibitive and also leads to considerable latency, timestamp perturbations, and data loss~\cite{gehrig2022arxiv}. As such, current research endeavors are in favor of offline, plug-in-play neuromorphic super-resolution (\textbf{SR}) algorithms. Specifically, event-level SR is a process of enhancing a low-resolution (\textbf{LR}) source to its high-resolution (\textbf{HR}) counterpart\footnote{For clarity, HR estimates from a LR source by external algorithms are called SR events.}. It simultaneously handles spatial and temporal information to shape a new stream of four-dimensional (4D) events, distinguishing itself from frame-based SR that simply operates on a 2D plane. In regard to benefits derived, SR events can deliver enriched visualization of dynamic scenes, performance gains in event-driven analysis, along with much more potential for synergistic integration with other imaging modalities~\cite{gehrig2022arxiv,li2019neurocomputing}.

\begin{table*}[t]
\caption{Evaluations of existing neuromorphic SR methods.}
\label{table:intro_comp}
\centering
\begin{tabular}{lccccc}
    \toprule
    \textbf{Method} & Learning-based & Free of auxiliary data & Free of prior training & Spatiotemporal SR & Large-scale ($8\times$, $16\times$) SR\\ \midrule
    Li~\etal~\cite{li2019neurocomputing} & \xmark & \cmark & \cmark & \xmark & \xmark\\
    Wang~\etal~\cite{wang2020cvpr} & \xmark & \xmark & \cmark & \xmark & \cmark\\
    Duan~\etal~\cite{duan2021cvpr} & \cmark & \xmark & \xmark & \xmark & \xmark\\
    Li~\etal~\cite{li2021iccv} & \cmark & \xmark & \xmark & \cmark & \xmark\\ 
    Weng~\etal~\cite{weng2022eccv} & \cmark & \xmark & \xmark & \xmark & \cmark\\ 
    Huang~\etal~\cite{huang2024cvpr} & \cmark & \xmark & \xmark & \xmark & \xmark\\ \midrule
    Ours & \cmark &  \cmark&  \cmark & \cmark & \cmark\\
    \bottomrule
\end{tabular}
\end{table*}
So far, there have already been several investigations unleashing the capability of events from LR constraints~\cite{li2019neurocomputing,wang2020cvpr,duan2021cvpr,li2021iccv,weng2022eccv,huang2024cvpr}, and~\Cref{table:intro_comp} presents evaluations from the following aspects:
\begin{enumerate}
    \item Learning-based techniques have showed their superiority over model-based ones in both reconstruction quality and downstream applications~\cite{duan2021cvpr,li2021iccv,weng2022eccv,huang2024cvpr}.
    
    \item Supervised fashions are restricted by real data availability. Not only are current neuromorphic cameras unable to perfectly support high spatial resolution~\cite{gehrig2022arxiv}, but collecting a huge volume of events is also laborious and even unfeasible since motion has the nature of randomness and variability. While we might require only one HR frame as a known fact for supervision, the result also submits to the quality of frame imaging~\cite{wang2020cvpr}.

    \item Exhaustive prior training imposes a significant burden in terms of computing power, data repositories, and lengthy periods of time. When trained on synthetic samples from a simulator with fixed acquisition settings, the models are prone to have biases and poor generalization to rare instances (\eg, micrographs) with diverse acquisition conditions and parameters~\cite{weng2022eccv}.

    \item Common practices convert events to a grid~\cite{li2019neurocomputing,wang2020cvpr,duan2021cvpr,weng2022eccv,huang2024cvpr}, where the temporal dimension is flattened, to execute spatial SR first. With time redistribution, events are then reshaped from a lower-dimensional SR grid. This stepwise manner, where temporal and spatial operations are weakly bound, fails to perform spatiotemporal SR and might induce timestamp perturbations as well as large deviations in statistical properties between LR and SR events, resulting in compromised temporal accuracy~\cite{li2021iccv}.

    \item Large-scale (\eg, $8\times$, $16\times$) neuromorphic SR is explored in depth by few efforts due to training constraints and data unavailability. It reflects the upper bound of an algorithm and is more practical in most real-world scenarios.
\end{enumerate}

This work aims to tackle the above challenges and presents the following innovation and contributions:
\begin{enumerate}
    \item We introduce the first self-supervised learning prototype for neuromorphic SR. It demonstrates that internal learning, which accommodates the model itself to different configurations per input sample taken by a LR camera, is sufficiently representative to estimate an accurate SR correspondence of any scale without lengthy training on external knowledge. It is thus free from the restriction of real data availability and adaptive to diverse imaging settings. This method is realized via the synergy between the two distinct representations of a single event stream, with each conveying complementary space-time features. We show that the prototype, developed on a convolutional neural network (CNN) and a multilayer perceptron (MLP), already achieves satisfactory results that can be further improved by exploiting advanced modules.

    \item Since our approach is not subject to prior knowledge, it can theoretically reach any SR scale. Assessed on downstream tasks, it delivers comparable results against the state-of-the-arts on $2\times$, $4\times$, $8\times$, and $16\times$ scale samples, and also reaches up to a $32\times$ level that previous counterparts fail to achieve. In addition, it effectively recovers and enhances inferior natural images and optical micrographs taken under non-ideal imaging conditions, overcoming the challenges that are difficult to address by conventional frame-based techniques.

    \item Finally, we show the superiority of our SR data processed from a LR source over the direct output from a HR neuromorphic camera, via both qualitative and quantitative comparisons. Given the limitations of current HR devices and the ongoing debate on their use in sensing and vision, our solution offers a practical and flexible alternative.
\end{enumerate}

\section{Related Work}
\subsection{Neuromorphic Cameras}
\begin{figure}[t]
    \centering
    \includegraphics[width=0.48\textwidth]{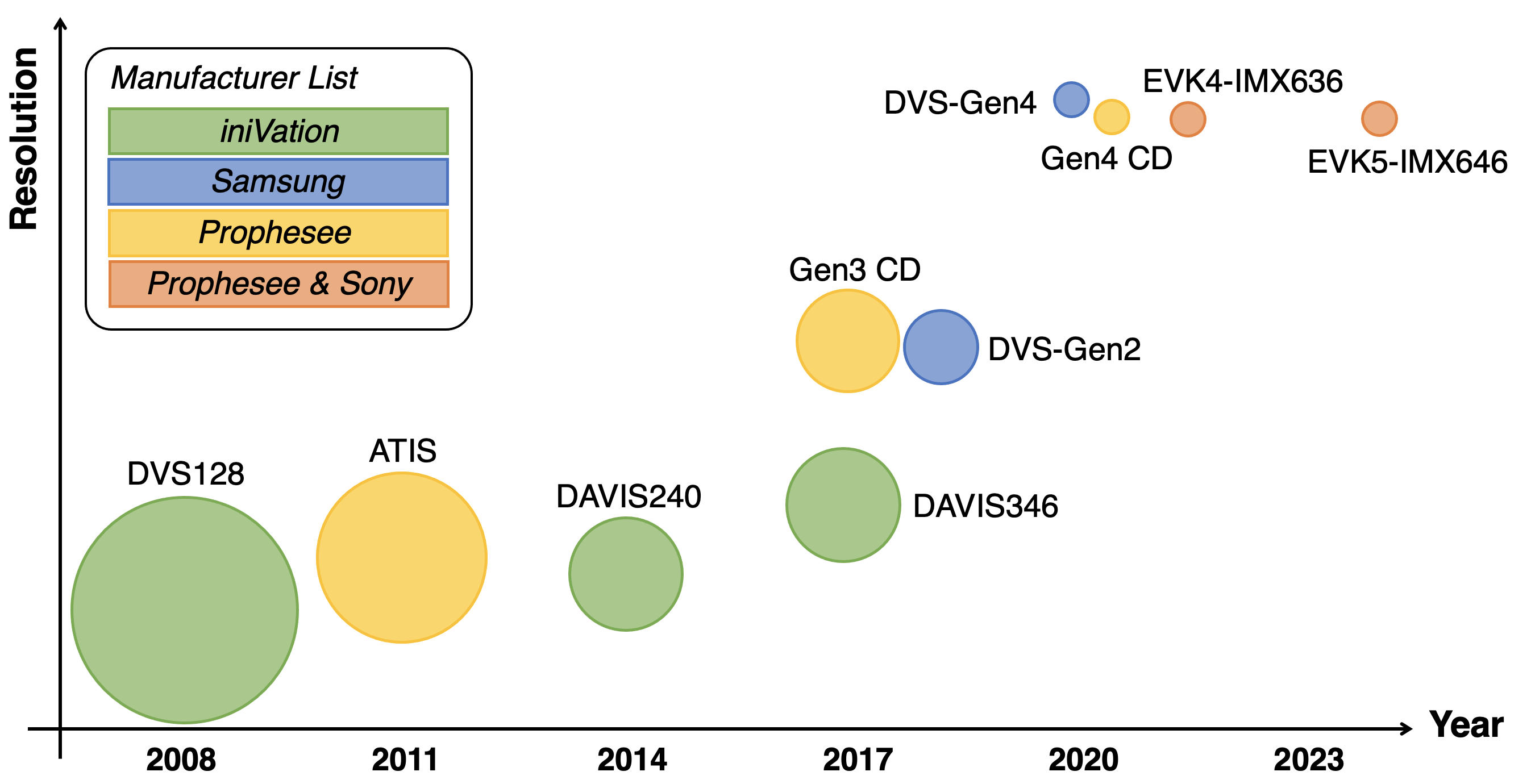}
    \caption{Development of neuromorphic cameras. The circle diameter stands for pixel size (in \si{\um}). Leading manufacturers are listed: \textit{iniVation}~\cite{lichtsteiner2008jssc,brandli2014jssc,taverni2018tcs}, \textit{Samsung}~\cite{son2017isscc,suh2020iscas}, \textit{Prophesee}~\cite{posch2010jssc,gen3cd,finateu2020isscc}, and \textit{Prophesee \& Sony}~\cite{evk4,evk5}.}
    \label{fig:cam_dev}
\end{figure}
Neuroscience research reveals that the human visual system interprets information in a hierarchical manner, with each layer of the retina performing a distinct function in visual perception~\cite{posch2014retinomorphic}. The dynamic vision sensor (iniVation, DVS128, $128 \times 128$ pixels)~\cite{lichtsteiner2008jssc}, born in $2008$, emulates a simplified three-layer retinal structure and mimics message flows in between. Bridging a DVS and an active pixel sensor (APS) in the same pixel, neuromorphic cameras can also support both frame-free and frame-based visual data output in parallel. As shown in~\cref{fig:cam_dev}, the market has seen a growing number of commercial products in recent years, with an advance typically on a three-year cycle. A common trend across manufacturers is to decrease pixel size for higher spatial resolution. Specifically, pixel size has been reduced from \SI{40}{\um} to as small as \SI{4.86}{\um}, improving resolution from $128 \times 128$ to $1280 \times 720$ pixels. This facilitates the mass production of sensors with larger pixel arrays and delivers a higher degree of fidelity and clarity for observational purposes. However, the event rate also rises as the increasing resolution, which places more burdens on bandwidth and computing requirements. Most HR cameras are facing two challenges --- limited readout rates and bus saturation, which often cause a higher noise level, timestamp perturbations, data loss, and accordingly compromised performance~\cite{gehrig2022arxiv}. As such, their use in imaging still remains a subject of debate.

\subsection{Frame Super-Resolution}
Conventional SR aims to upscale the spatial resolution of frames, and learning-based methods, which model a parametric transformation from LR sources to their HR correspondences, have showed dominance in this task~\cite{zhang2021tcsvt,zhao2021ai}. Increasing the network depth or width can obtain a dramatically elevated performance~\cite{huang2024tcsvt}, and another expectation is to pursue low computing costs while possessing high precision~\cite{zhu2022tcsvt}. These supervised approaches infer a HR estimate for an unseen frame from the knowledge of seen LR-HR pairs. To alleviate the reliance on large diverse datasets and lengthy training time, self-supervised learning is applied to frame SR. For example, Shocher~\etal~\cite{shocher2018cvpr} proposed a classic framework that is trained at the inference stage on samples derived from a LR frame itself. It can accomplish faithful HR estimations without auxiliary data support, which is particularly valuable when handling real-world or historic images whose ground-truth resources are often unavailable. Similarly, due to the deficiency of current HR neuromorphic cameras as well as the random, diverse nature of motion, it is also challenging to collect a huge quantity of high-quality events as a standard benchmark. In other words, each event stream is a unique and rare sample. With only LR events, we thus resort to self-supervision and internal learning in neuromorphic SR tasks.

\subsection{Neuromorphic Super-Resolution}
A few approaches have been developed for getting rid of LR restrictions on events. Li~\etal~\cite{li2019neurocomputing} presented the first model-based simulation of HR streaming events with a non-homogeneous Poisson process. GEF~\cite{wang2020cvpr} bridges two imaging modalities via motion compensation. By optimizing the joint contrast between the two sources, it can upscale a LR event frame to a HR image resolution, and then an optical-flow-based redistribution reshapes events from an estimated HR event frame~\cite{duan2022tpami}. Duan~\etal~\cite{duan2021cvpr} released the learning-based EventZoom, with an optimized event-to-image module, to learn a mapping from LR event stacks to HR ones. They also found that timestamp assignment schemes marginally influence the retrieval of events from lower-dimensional stacks. A supervised spatiotemporal constraint learning fashion based on an end-to-end spiking neural network is capable of estimating the space-time distribution in parallel~\cite{li2021iccv}, and the one using a recurrent network can achieve impressive large-factor $16 \times$ SR results~\cite{weng2022eccv}. A recently proposed bilateral network, where shared information in positive/negative events are fully leveraged, obtains a significant improvement~\cite{huang2024cvpr}. Regrettably, as~\Cref{table:intro_comp} shows, the above methods are subject to certain constraints. This work presents a new solution to spatiotemporally super-resolve neuromorphic event streams.

\section{Methodology}
\subsection{Problem Definition}
A neuromorphic camera with $H \times W$ spatial resolution generates streaming events in response to brightness changes from a moving target, with each event being denoted by
\begin{equation}\label{eq:ev}
    \mathbf{e}_i = (\mathbf{x}, t_i, p_i),
\end{equation}
in which an event $\mathbf{e}_i$, indexed by $i$, is triggered at time $t_i \leq T$ in a pixel $\mathbf{x}$, and $T$ is the timestamp at which the last event is triggered. The position $\mathbf{x} = (x, y)^{\mathsf{T}}$ consists of two orthogonal directions $x \in \interval[scaled]{1}{W}$ and $y \in \interval[scaled]{1}{H}$. The binary polarity $p_i \in \{-1, +1\}$ represents the sign of the brightness change. Then, a complete stream $\mathbf{E}$ is
\begin{equation}
    \mathbf{E} = \{\mathbf{e}_i\}_{i=1:\infty} = \big\{(\mathbf{x}, p_i, t_i)\big\}_{i=1:\infty}.
\end{equation}
Given a LR input source $\mathbf{E}$, we infer its SR counterpart 
\begin{equation}
    \hat{\mathbf{E}} = f(\mathbf{E}, \sigma)
\end{equation}
such that each event $\mathbf{e}_i = (\hat{\mathbf{x}}, \hat{t}_i, \hat{p}_i) \in \hat{\mathbf{E}}$ has $\hat{x} \in \interval[scaled]{1}{\sigma W}$, $\hat{y} \in \interval[scaled]{1}{\sigma H}$, $\hat{t}_i \leq T$, and $\hat{p}_i \in \{-1, +1\}$, where $\sigma$ is a scaling factor. We elaborate the implementation details of the proposed workflow $f$ in what follows.

\begin{figure*}[t]
    \centering
    \frame{\includegraphics[width = \textwidth]{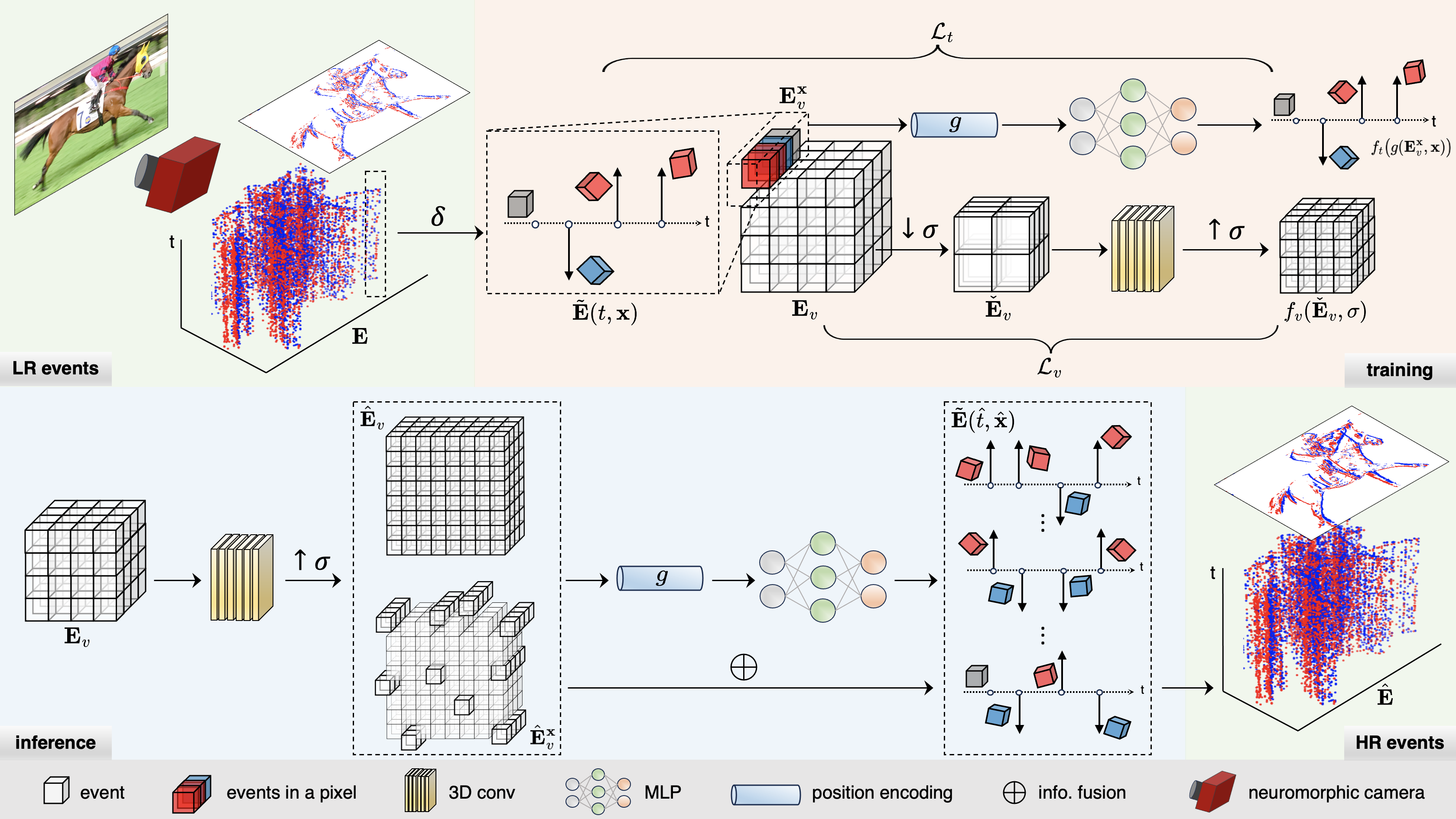}}
    \caption{Overview of our self-supervised neuromorphic SR prototype. The neural networks in two branches are trained at inference time. Best viewed in color.}
    \label{fig:workflow}
\end{figure*}
\subsection{Self-Supervised Workflow}
\cref{fig:workflow} depicts the prototype of our self-supervised neuromorphic SR, which consists of one LR input, two processing stages where the model is trained at test time, and one SR output. A camera captures a moving target and generates streaming LR events. In addition to~\cref{eq:ev} that takes an event as a space-time node, it can also be modelled as an impulse due to the continuously-varying time  
\begin{equation}\label{eq:ev_}
    \tilde{\mathbf{e}}_i(t, \mathbf{x}) = p_i^{\mathbf{x}} c \delta(t - t_i^{\mathbf{x}}),
\end{equation}
where $\delta(t)$ is the Dirac delta function, and $c$ is the contrast threshold. \cref{eq:ev_} applies to every pixel in the coordinate. Then, a stream in a specific $\mathbf{x}$ has an impulse-train representation
\begin{equation}
    \tilde{\mathbf{E}}(t, \mathbf{x}) = \sum_{i=1}^{\infty} \tilde{\mathbf{e}}_i(t, \mathbf{x}) = \sum_{i=1}^{\infty} p_i^{\mathbf{x}} c \delta(t - t_i^{\mathbf{x}}),
\end{equation}
which precisely encodes timestamp information for each event.

The self-supervised spatiotemporal operation requires a new representation --- event voxel-grid, with each voxel featured by $p_i$ representing that an event exists in a particular position. The voxel-grid is a 3D tensor $\mathbf{E}_v \in \mathbb{R}^{L \times H \times W}$ where $L$ denotes the upper bound of the length $L^{\mathbf{x}}$ of an event stream among all positions $\mathbf{X}$, with
\begin{align}
    L^{\mathbf{x}} &= \frac{1}{c}\sum_{i=1}^{\infty} \bigg|~p_i^{\mathbf{x}} c \int_{0}^{T}\delta(t - t_i^{\mathbf{x}})~\mathrm{d}t~\bigg|,\label{eq:lx}\\
    L &= \max \left\{L^{\mathbf{x}} \mid \mathbf{x} \in \mathbf{X}\right\}.
\end{align}
Zero-padding fills the voxel if $L^{\mathbf{x}} < L$. However, such a definition might induce sparsity in the tensor that could negatively affect computations. To improve model performance, we scale the features to an appropriate range and offset the zero-padding to make the tensor denser. Therefore, a column of voxels $\mathbf{E}_v^{\mathbf{x}} \in \mathbb{R}^{L \times 1 \times 1}$ of a certain $\mathbf{x}$, which encodes the polarity and the size of a stream, associates with $\tilde{\mathbf{E}}(t, \mathbf{x})$ that records the corresponding microsecond timestamps. 

In contrast to previous research in which a voxel describes a time bin that leads to a significant reduction in the original temporal resolution~\cite{zhu2019cvpr}, ours thoroughly discards the time dimension and only represents the position, polarity, and event quantity of a scene record. As such, the proposed workflow harnesses two event-based manifestations to supply complementary information, resulting in two branches for handling the received samples.

\begin{figure}[t]
    \centering
    \subfloat[Image]{\frame{\includegraphics[width = 0.23\textwidth]{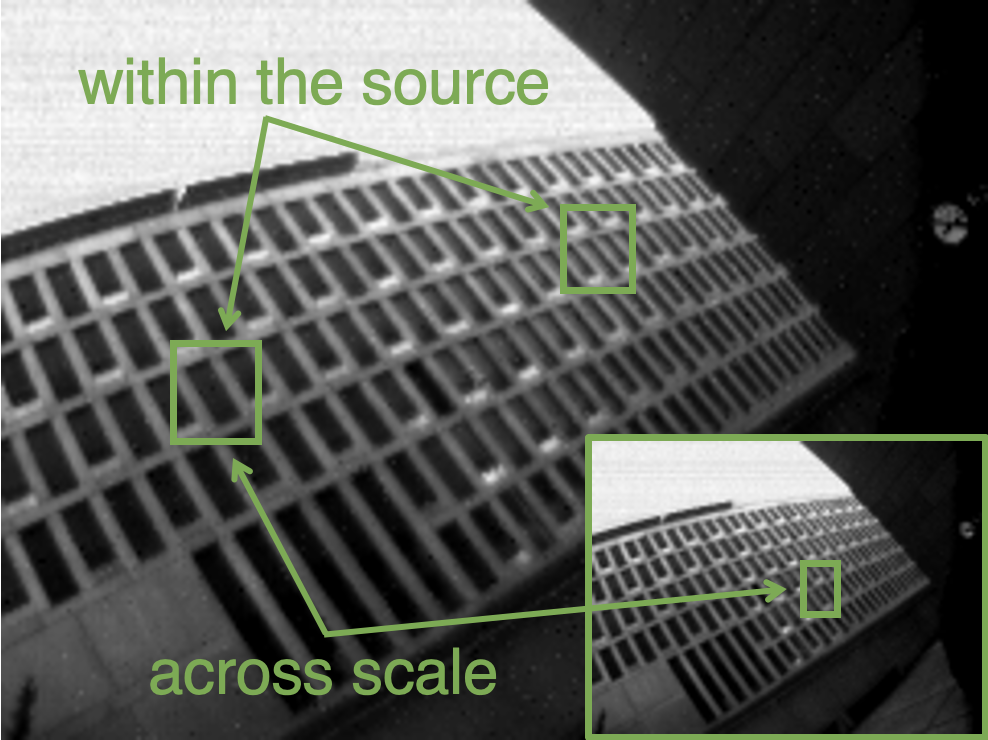}}}\hfill
    \subfloat[2D visualization of an event voxel-grid]{\frame{\includegraphics[width = 0.23\textwidth]{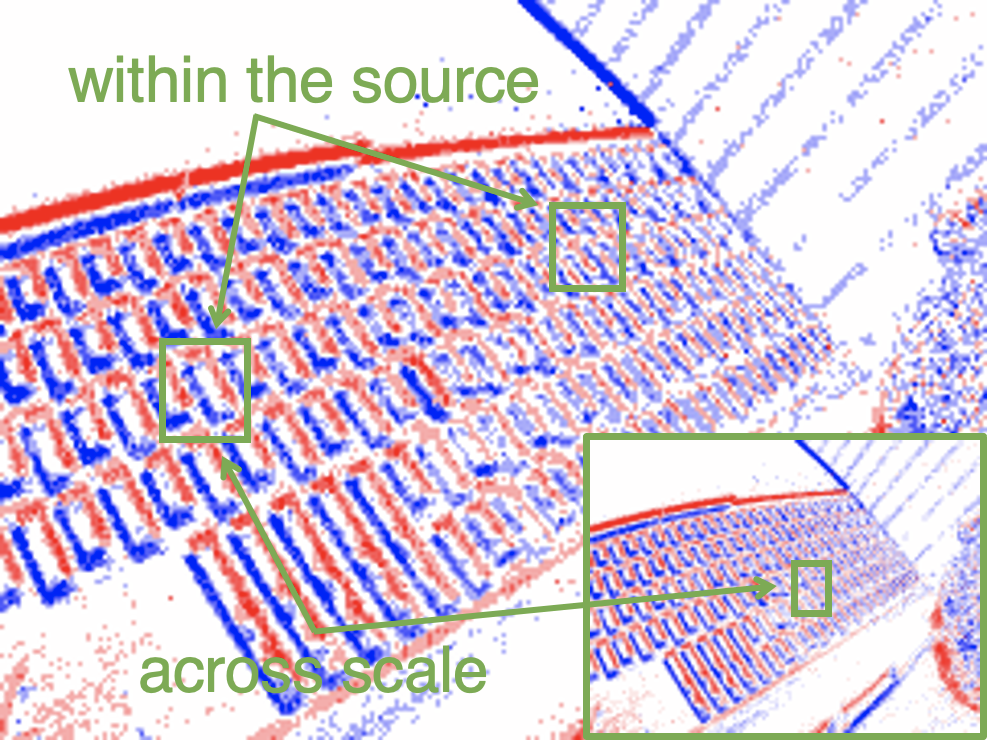}}}
    \caption{Spatially, small patches often exhibit a strong degree of recurrence within the source and across its coarser scales. Image courtesy of~\cite{mueggler2017ijrr}.}
    \label{fig:demo_spatial}
\end{figure}
\textbf{Spatial Dimension.} Previous explorations empirically verified that there is substantial recurrence of information inside a single image of low visual entropy, and it is thus possible to estimate any SR counterparts for the image by simply observing its internal statistics of strong predictive-power~\cite{zontak2011cvpr,shocher2018cvpr,zhao2023icip}. Neuromorphic imaging fails to capture low-frequency components and produces much simpler structures of scenes, leading to a considerable increase in data repetition and redundancy in an event-based grid itself. An example is shown in~\cref{fig:demo_spatial}, where patch repetition exists in the source and across its other scales, for both the image and event data. Apparently, the scene already has adequate recurrence of a region of interest (ROI), in different positions at different scales. This is magnified in the events due to the missing absolute intensity. Thus, it is possible to learn a mapping between a LR source and its SR reconstruction in an internal way.

We bicubically downsample the LR input $\mathbf{E}_v$ with $\sigma$ to acquire its coarser resolution $\check{\mathbf{E}}_v$. Then, a 3D CNN $f_v$ transforms $\check{\mathbf{E}}_v$ to $\mathbf{E}_v$ via a loss function
\begin{equation}\label{eq:downsample}
    \mathcal{L}_v = \left\lVert f_v(\check{\mathbf{E}}_v, \sigma) - \mathbf{E}_v \right\rVert_1.
\end{equation}
As the only one instance $\mathbf{E}_v$ is not sufficiently representative, data augmentation is exploited to enrich more LR-HR pairs in training. Following the common practice~\cite{shocher2018cvpr}, three directions of rotation (\SI{90}{\degree}, \SI{180}{\degree}, and \SI{270}{\degree}), as well as horizontal and vertical mirror reflections are made to each pair. Lastly, the well-trained $f_v$, which leverages data inherited similarity, infers a SR version upon the LR input
\begin{equation}
    \hat{\mathbf{E}}_v = f_v(\mathbf{E}_v, \sigma).
\end{equation}

\begin{figure}[t]
    \centering
    \includegraphics[width = 0.45\textwidth]{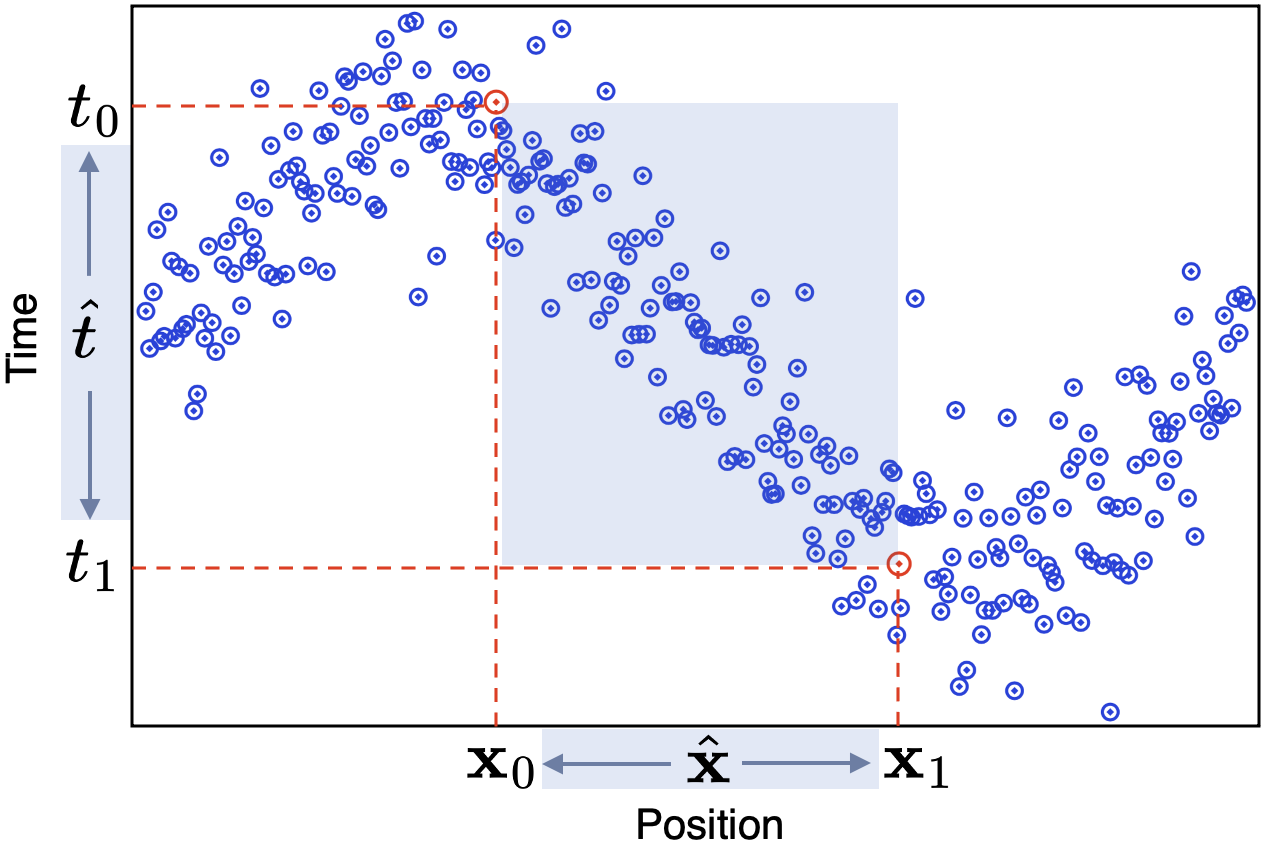}
    \caption{Distribution of a SR event stream in terms of spatial positions and timestamps. In the same normalized coordinate, red pixels stand for the LR events where $\mathbf{x} = \hat{\mathbf{x}}$, and blue subpixels mark the SR events for which $f_t$ trained on LR distribution (\ie, $\mathbf{x}_0$, $\mathbf{x}_1$) makes timestamp estimations.}
    \label{fig:demo_time}
\end{figure}

\textbf{Temporal Dimension.} Owing to the randomness and variability of motion, each event stream has a unique sequence of timestamps, which cannot be found in any external database. Estimating microsecond timestamps for new SR events can be hypothesized as a regression task accomplished by a MLP. As described above, each $\mathbf{E}_v^{\mathbf{x}}$ drops explicit time information that the associated $\tilde{\mathbf{E}}(t, \mathbf{x})$ preserves. We harness a MLP $f_t$ to model the relationship among spatial positions, polarity, and timestamps, with a loss function
\begin{equation}\label{eq:loss_t}
    \mathcal{L}_t = \frac{1}{H W}\sum^{H W}_{\mathbf{x}} \int_{0}^{T} \left(f_t\big(g(\mathbf{E}_v^{\mathbf{x}}, \mathbf{x})\big) - \tilde{\mathbf{E}}(\tau, \mathbf{x})\right)^2~\mathrm{d}\tau,
\end{equation}
where a hard-rule encoder $g$ bridges $\mathbf{E}_v^{\mathbf{x}}$ with its explicit position $\mathbf{x}$. To simplify computations, we directly stack the two features $\mathbf{E}_v^{\mathbf{x}}$ and $\mathbf{x}$ in a single vector. As such, $g(\mathbf{E}_v^{\mathbf{x}}, \mathbf{x})$ is a result containing both the spatial position and polarity, and it requires a mapping to the corresponding timestamps stored in $\tilde{\mathbf{E}}(t, \mathbf{x})$. $\hat{\mathbf{E}}_v$ is an upscale of $\mathbf{E}_v$ where each pixel is expanded by a factor of $\sigma$ along both axes. Our target is to estimate the timestamps of the new events in the expanded pixels. As the spatiotemporal correlation principle claims, an event strongly correlates with its neighbors in space-time, giving that any two spatially-adjacent events should share close timestamps~\cite{chen2023tcsvt}. Then, SR events have an impulse-train $\tilde{\mathbf{E}}(\hat{t}, \hat{\mathbf{x}})$, obtained by
\begin{equation}
    \tilde{\mathbf{E}}(\hat{t}, \hat{\mathbf{x}}) = f_t\big(g(\hat{\mathbf{E}}_v^{\mathbf{x}}, \hat{\mathbf{x}})\big).
\end{equation}
If the computations are on the same normalized coordinate, there is $\mathbf{E}_v \subseteq \hat{\mathbf{E}}_v$ for a well-trained $f_v$ that has exploited spatial pixel repetition. Similarly, $f_t$, which has observed the mapping in the LR source by~\cref{eq:loss_t}, predicts the timestamps $\hat{t}$ of the SR events in the generated subpixels as per the inherited similarity. \cref{fig:demo_time} presents an illustration. In inference, a fine-tuned $f_t$ can memorize the correct $\hat{t}$, which equals to $t$ in $\hat{\mathbf{x}} = \mathbf{x}$. Thus, neuromorphic SR in the temporal dimension can be simplified as a regression task that makes a prediction for a SR subpixel input $\hat{\mathbf{x}}$, where $\lVert\mathbf{x}_0\rVert < \lVert\hat{\mathbf{x}}\rVert < \lVert\mathbf{x}_1\rVert$, and $\mathbf{x}_0$ is adjacent to $\mathbf{x}_1$ in the LR source. We exploit $f_t$, which has learned the mapping from $(\mathbf{x}_0, \mathbf{x}_1)$ to $(t_0, t_1)$, to estimate $\hat{t}$ from $\hat{\mathbf{x}}$. The resulting $\hat{t}$ of $\hat{\mathbf{x}}$ closely follows $t_0$ of $\mathbf{x}_0$ since $f_t$ is trained on a LR source obeying the spatiotemporal correlation. 

\textbf{Overall Workflow.} Integrating both the spatial and temporal branches, the workflow of our self-supervised method shown in~\cref{fig:workflow} is described as follows. An event sample $\mathbf{E}$ captured by a LR neuromorphic camera is modelled as an event voxel-grid $\mathbf{E}_v$. We train a spatial mapping from $\mathbf{E}_v$ to its coarser resolution $\check{\mathbf{E}}_v$ via a 3D CNN $f_v$ with a loss function $\mathcal{L}_v$, which then makes inference from $\mathbf{E}_v$ to its spatial SR estimate $\hat{\mathbf{E}}_v$. Meanwhile, we model the relationship among spatial positions, polarity, and timestamps within an event stream, where a MLP $f_t$ with a loss function $\mathcal{L}_t$ learns a transformation from $\mathbf{E}_v^{\mathbf{x}}$ to the corresponding timestamps $\tilde{\mathbf{E}}(\tau, \mathbf{x})$. Then, the trained $f_t$ estimates new timestamps $\tilde{\mathbf{E}}(\hat{t}, \hat{\mathbf{x}})$ for $\hat{\mathbf{E}}_v$. Finally, we integrate both the spatial and temporal SR features to obtain a complete SR event stream $\hat{\mathbf{E}}$. There are thus only three elements in our approach --- one LR input, one self-driven pipeline, and one SR output.

\newcommand\wh{80pt}
\newcommand\whs{44pt}
\newcommand\hs{-22pt}
\newcommand\shs{-8pt}
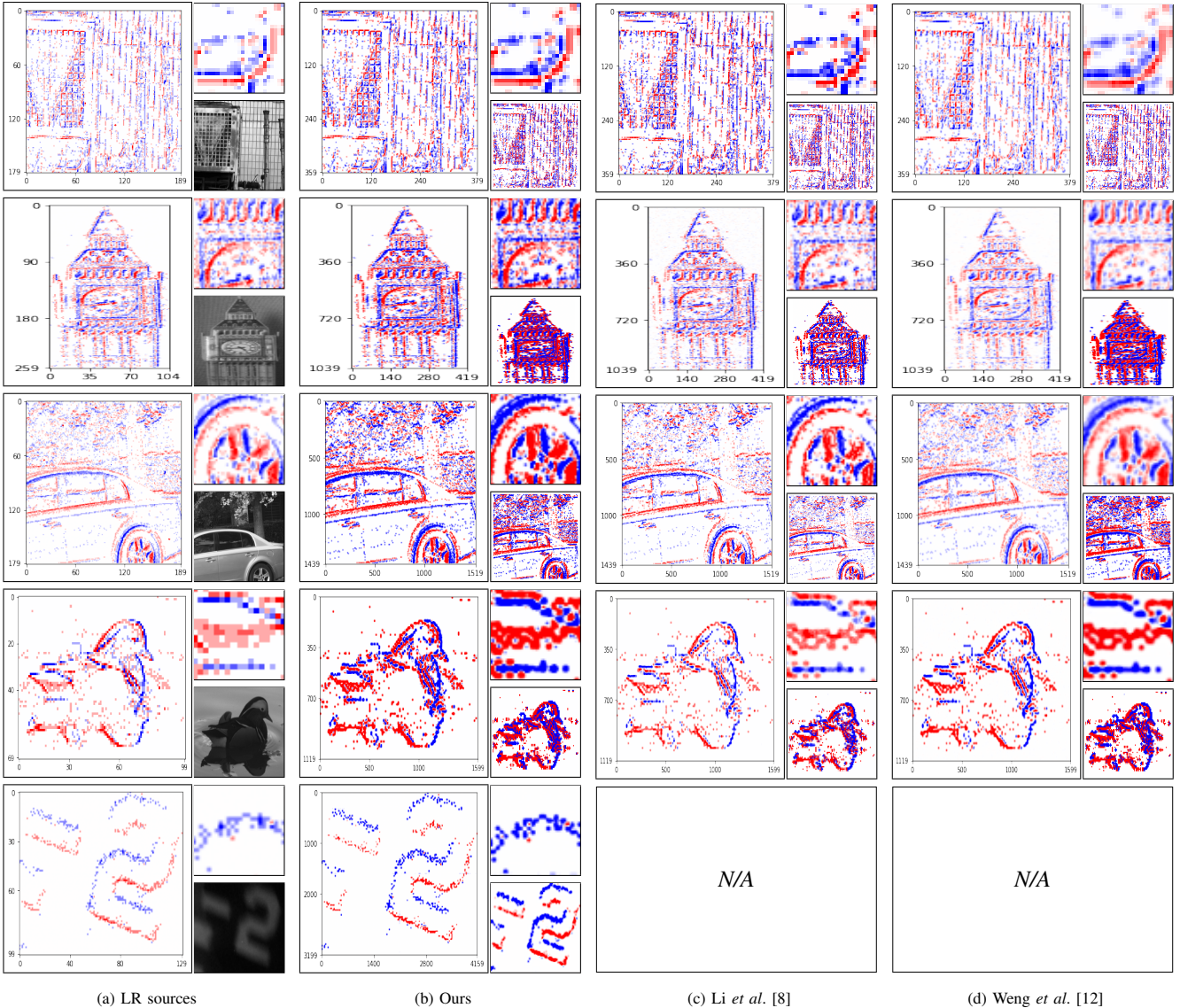
\begin{figure*}[p]
    \centering
    \hspace{-10pt}
    \subfloat[LR sources]{
    \begin{tabular}[b]{c}
        \frame{\includegraphics[width=92pt, height=\wh]{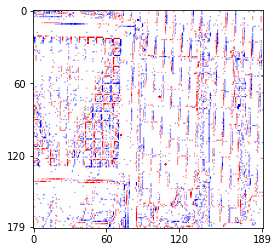}}\\[-2pt]        
        \begin{tabular}[b]{c}
            \frame{\includegraphics[width=\whs, height=\whs]{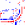}}
            \frame{\includegraphics[width=\whs, height=\whs]{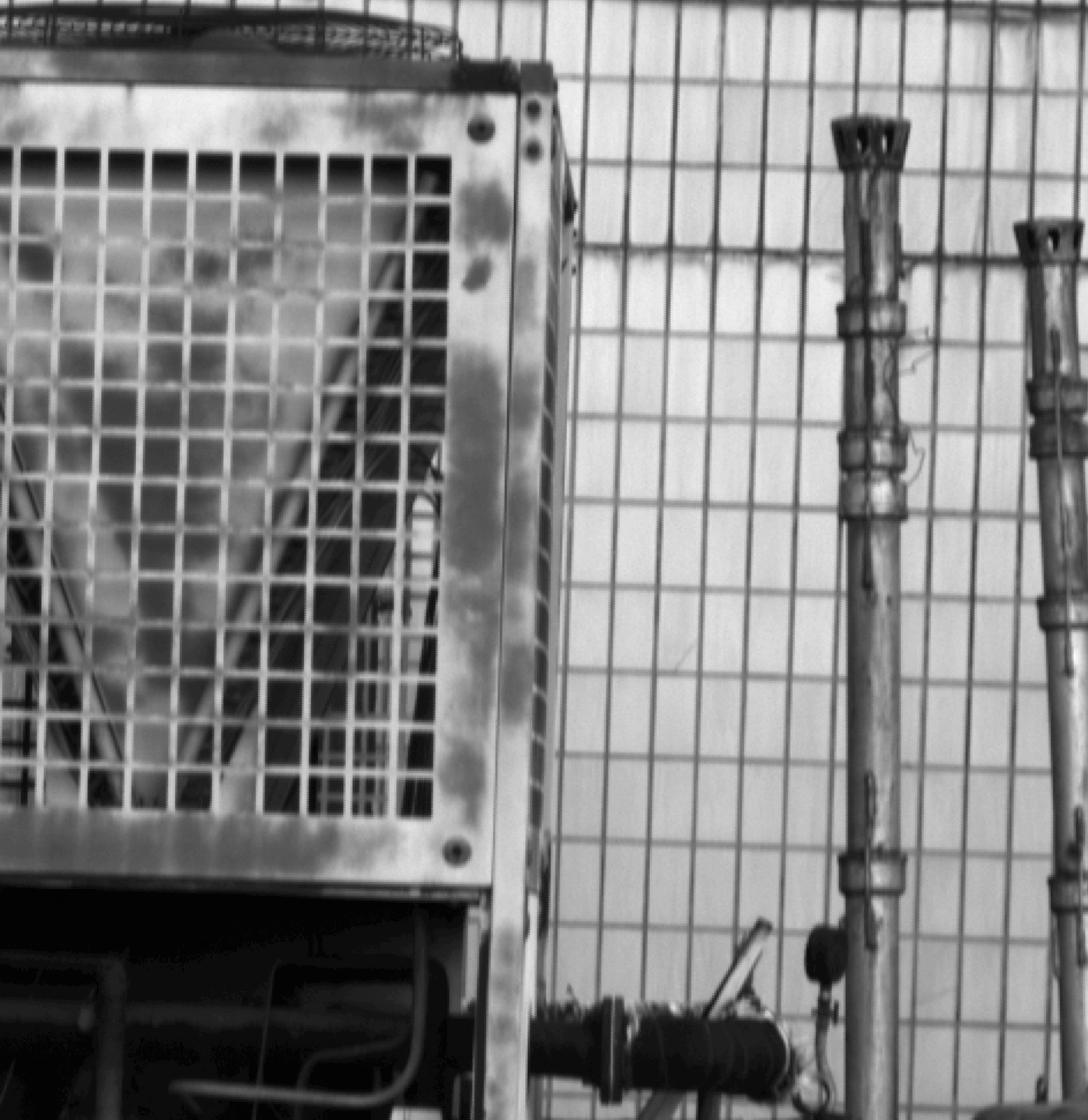}}
        \end{tabular}\\
        \frame{\includegraphics[width=92pt, height=\wh]{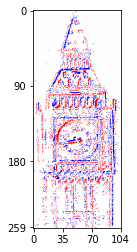}}\\[-2pt] 
        \begin{tabular}[b]{c}
            \frame{\includegraphics[width=\whs, height=\whs]{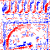}}
            \frame{\includegraphics[width=\whs, height=\whs]{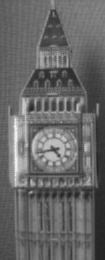}}
        \end{tabular}\\
        \frame{\includegraphics[width=92pt, height=\wh]{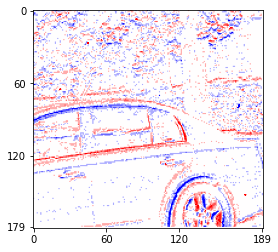}}\\[-2pt] 
        \begin{tabular}[b]{c}
            \frame{\includegraphics[width=\whs, height=\whs]{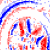}}
            \frame{\includegraphics[width=\whs, height=\whs]{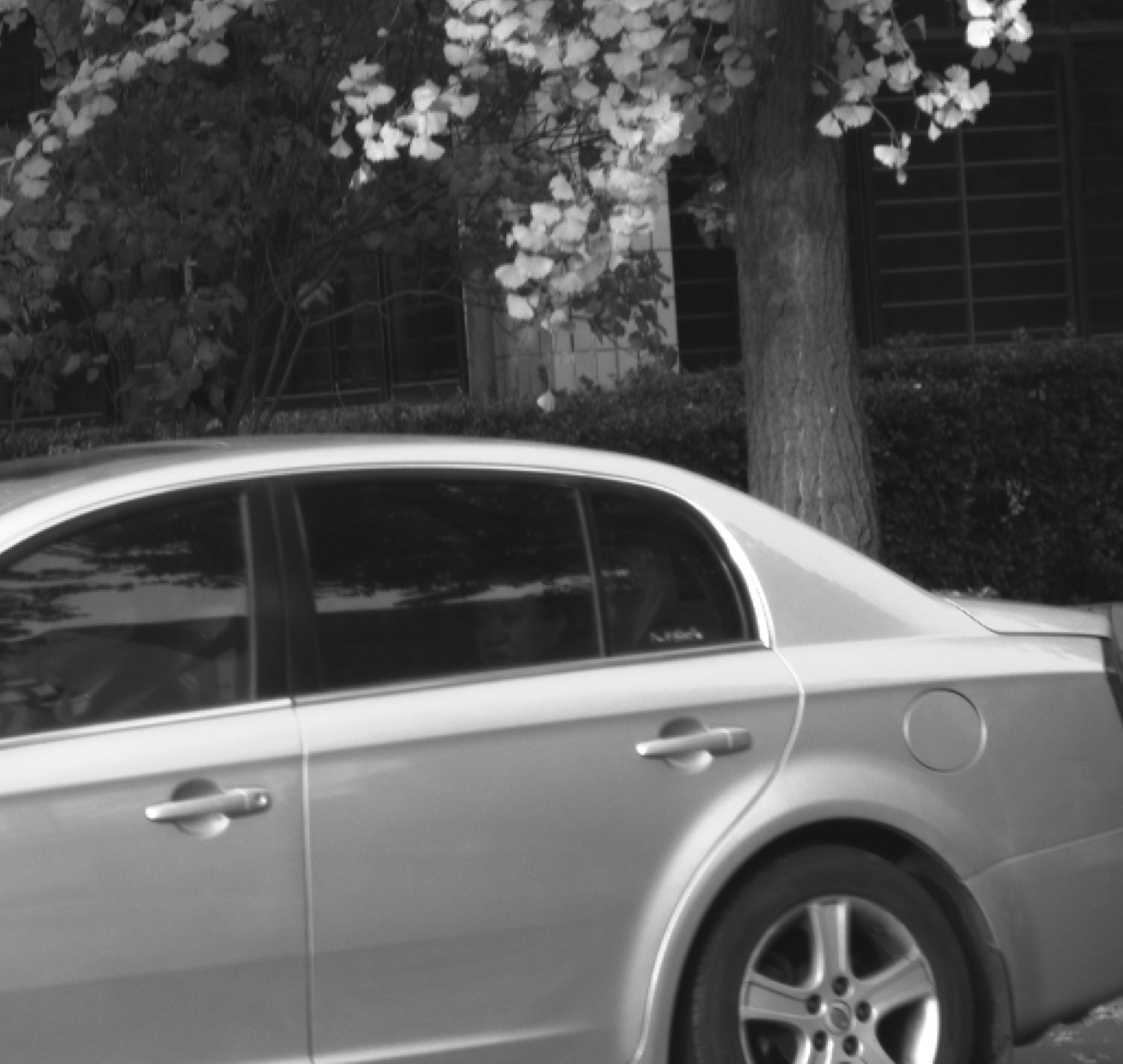}} 
        \end{tabular}\\
        \frame{\includegraphics[width=92pt, height=\wh]{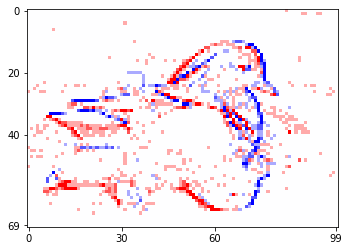}}\\[-2pt] 
        \begin{tabular}[b]{c}
            \frame{\includegraphics[width=\whs, height=\whs]{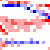}}
            \frame{\includegraphics[width=\whs, height=\whs]{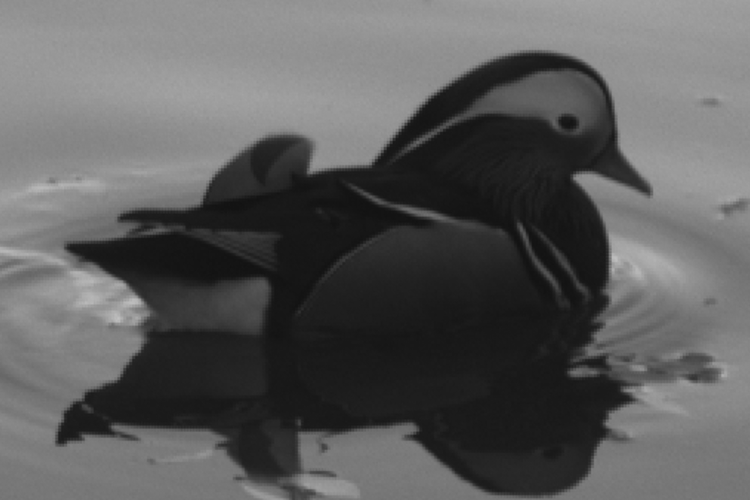}}
        \end{tabular}\\
        \frame{\includegraphics[width=92pt, height=\wh]{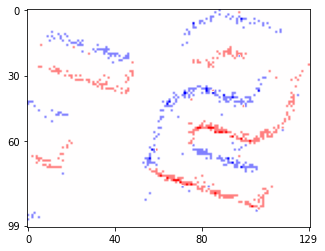}}\\[-2pt] 
        \begin{tabular}[b]{c}
            \frame{\includegraphics[width=\whs, height=\whs]{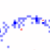}}
            \frame{\includegraphics[width=\whs, height=\whs]{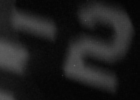}}
        \end{tabular}
    \end{tabular}}\hspace{\hs}
    \subfloat[Ours]{
    \begin{tabular}[b]{c}
        \frame{\includegraphics[width=92pt, height=\wh]{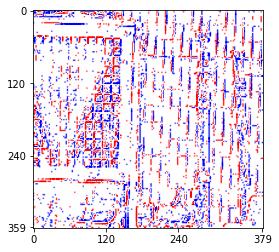}}\\[-2pt]        
        \begin{tabular}[b]{c}
            \frame{\includegraphics[width=\whs, height=\whs]{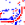}}
            \frame{\includegraphics[width=\whs, height=\whs]{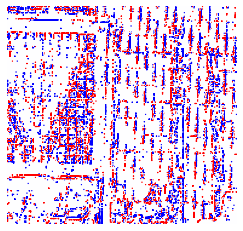}}
        \end{tabular}\\
        \frame{\includegraphics[width=92pt, height=\wh]{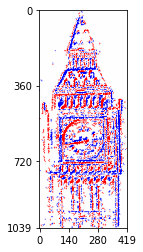}}\\[-2pt]  
        \begin{tabular}[b]{c}
            \frame{\includegraphics[width=\whs, height=\whs]{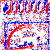}}
            \frame{\includegraphics[width=\whs, height=\whs]{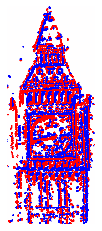}} 
        \end{tabular}\\
        \frame{\includegraphics[width=92pt, height=\wh]{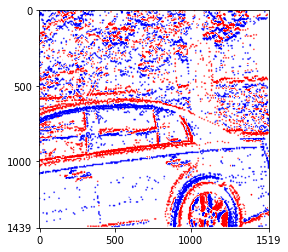}}\\[-2pt] 
        \begin{tabular}[b]{c}
            \frame{\includegraphics[width=\whs, height=\whs]{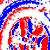}}
            \frame{\includegraphics[width=\whs, height=\whs]{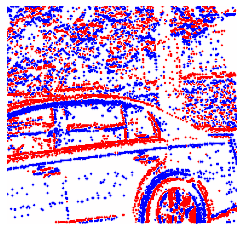}}
        \end{tabular}\\
        \frame{\includegraphics[width=92pt, height=\wh]{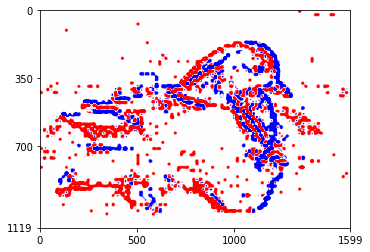}}\\[-2pt] 
        \begin{tabular}[b]{c}
            \frame{\includegraphics[width=\whs, height=\whs]{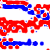}}
            \frame{\includegraphics[width=\whs, height=\whs]{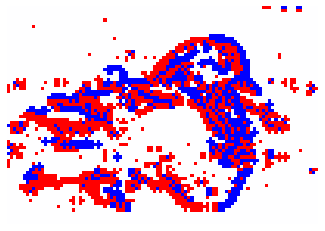}} 
        \end{tabular}\\
        \frame{\includegraphics[width=92pt, height=\wh]{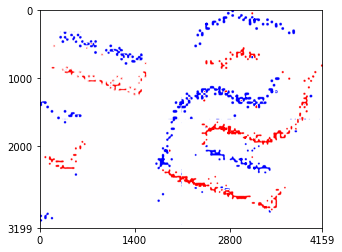}}\\[-2pt] 
        \begin{tabular}[b]{c}
            \frame{\includegraphics[width=\whs, height=\whs]{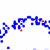}}
            \frame{\includegraphics[width=\whs, height=\whs]{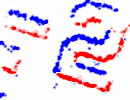}}
        \end{tabular}
    \end{tabular}}\hspace{\hs}
    \subfloat[Li~\etal~\cite{li2019neurocomputing}]{
    \begin{tabular}[b]{c}
        \frame{\includegraphics[width=92pt, height=\wh]{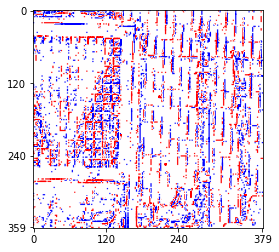}}\\[-2pt]        
        \begin{tabular}[b]{c}
            \frame{\includegraphics[width=\whs, height=\whs]{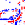}}
            \frame{\includegraphics[width=\whs, height=\whs]{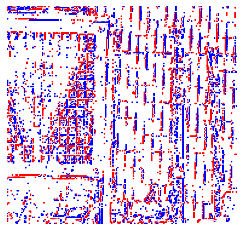}}
        \end{tabular}\\
        \frame{\includegraphics[width=92pt, height=\wh]{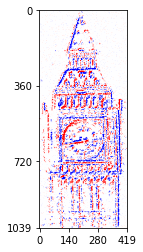}}\\[-2pt] 
        \begin{tabular}[b]{c}
            \frame{\includegraphics[width=\whs, height=\whs]{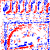}}
            \frame{\includegraphics[width=\whs, height=\whs]{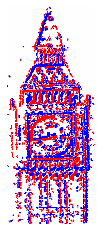}}
        \end{tabular}\\
        \frame{\includegraphics[width=92pt, height=\wh]{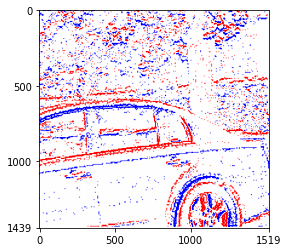}}\\[-2pt] 
        \begin{tabular}[b]{c}
            \frame{\includegraphics[width=\whs, height=\whs]{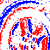}}
            \frame{\includegraphics[width=\whs, height=\whs]{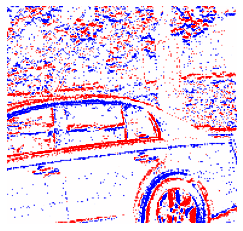}} 
        \end{tabular}\\
        \frame{\includegraphics[width=92pt, height=\wh]{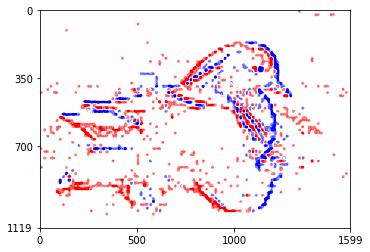}}\\[-2pt] 
        \begin{tabular}[b]{c}
            \frame{\includegraphics[width=\whs, height=\whs]{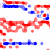}}
            \frame{\includegraphics[width=\whs, height=\whs]{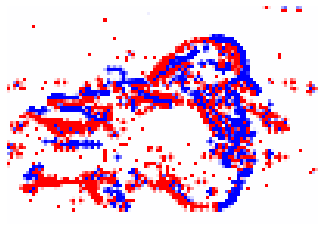}} 
        \end{tabular}\\
    \begin{tikzpicture}
        \draw[line width=0.2pt] (-1.6, 0) rectangle (1.6, 4.4);
        \node at (0, 2) {\textit{N/A}};
    \end{tikzpicture}
    \end{tabular}}\hspace{\hs}
    \subfloat[Duan~\etal~\cite{duan2021cvpr}]{
    \begin{tabular}[b]{c}
        \frame{\includegraphics[width=92pt, height=\wh]{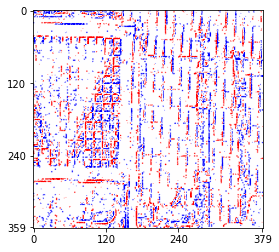}}\\[-2pt]        
        \begin{tabular}[b]{c}
            \frame{\includegraphics[width=\whs, height=\whs]{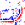}}
            \frame{\includegraphics[width=\whs, height=\whs]{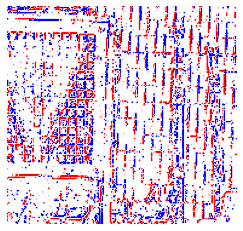}}
        \end{tabular}\\
        \frame{\includegraphics[width=92pt, height=\wh]{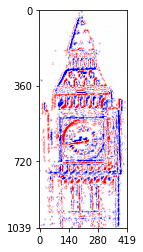}}\\[-2pt]  
        \begin{tabular}[b]{c}
            \frame{\includegraphics[width=\whs, height=\whs]{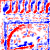}}
            \frame{\includegraphics[width=\whs, height=\whs]{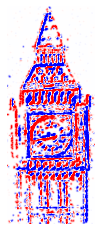}}
        \end{tabular}\\
        \frame{\includegraphics[width=92pt, height=\wh]{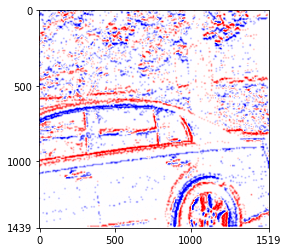}}\\[-2pt] 
        \begin{tabular}[b]{c}
            \frame{\includegraphics[width=\whs, height=\whs]{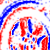}}
            \frame{\includegraphics[width=\whs, height=\whs]{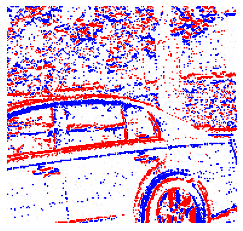}} 
        \end{tabular}\\
        \frame{\includegraphics[width=92pt, height=\wh]{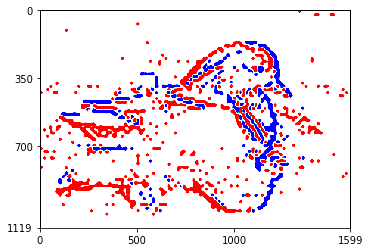}}\\[-2pt] 
        \begin{tabular}[b]{c}
            \frame{\includegraphics[width=\whs, height=\whs]{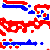}}
            \frame{\includegraphics[width=\whs, height=\whs]{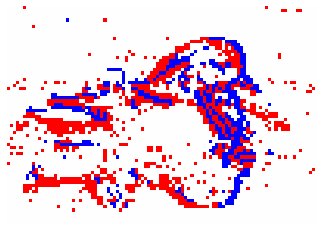}} 
        \end{tabular}\\
    \begin{tikzpicture}
        \draw[line width=0.2pt] (-1.6, 0) rectangle (1.6, 4.4);
        \node at (0, 2) {\textit{N/A}};
    \end{tikzpicture}
    \end{tabular}}\hspace{\hs}
    \subfloat[Weng~\etal~\cite{weng2022eccv}]{
    \begin{tabular}[b]{c}
        \frame{\includegraphics[width=92pt, height=\wh]{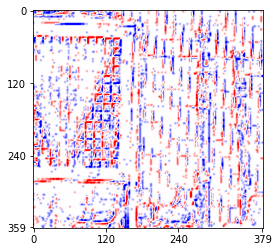}}\\[-2pt]         
        \begin{tabular}[b]{c}
            \frame{\includegraphics[width=\whs, height=\whs]{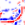}}
            \frame{\includegraphics[width=\whs, height=\whs]{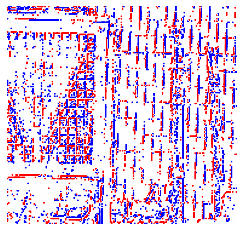}}
        \end{tabular}\\
        \frame{\includegraphics[width=92pt, height=\wh]{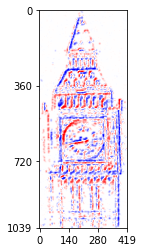}}\\[-2pt] 
        \begin{tabular}[b]{c}
            \frame{\includegraphics[width=\whs, height=\whs]{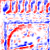}}
            \frame{\includegraphics[width=\whs, height=\whs]{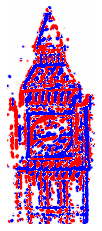}}
        \end{tabular}\\
        \frame{\includegraphics[width=92pt, height=\wh]{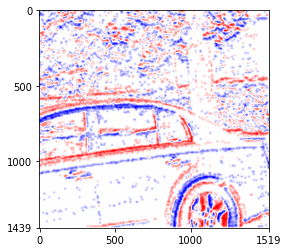}}\\[-2pt] 
        \begin{tabular}[b]{c}
            \frame{\includegraphics[width=\whs, height=\whs]{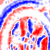}}
            \frame{\includegraphics[width=\whs, height=\whs]{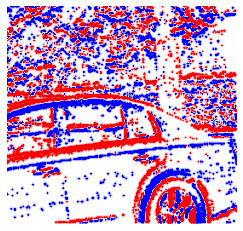}} 
        \end{tabular}\\
        \frame{\includegraphics[width=92pt, height=\wh]{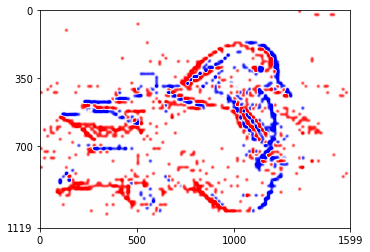}}\\[-2pt] 
        \begin{tabular}[b]{c}
            \frame{\includegraphics[width=\whs, height=\whs]{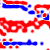}}
            \frame{\includegraphics[width=\whs, height=\whs]{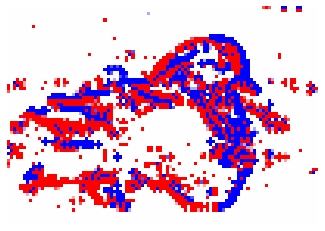}} 
        \end{tabular}\\
    \begin{tikzpicture}
        \draw[line width=0.2pt] (-1.6, 0) rectangle (1.6, 4.4);
        \node at (0, 2) {\textit{N/A}};
    \end{tikzpicture}
    \end{tabular}}
    \caption{From top to down, frame-based visualization of neuromorphic SR results of $2 \times$~\cite{wang2020cvpr}, $4 \times$~\cite{zhang2023tci}, $8 \times$~\cite{wang2020cvpr}, $16 \times$~\cite{wang2020cvpr}, and $32 \times$~\cite{zhang2024tip}. Results of (c) and (d) are not applicable in $32 \times$. For each sample, we also zoom in a certain ROI (lower left) and have a subpixel view in the original resolution (lower right).}
    \label{fig:vis_sr}
\end{figure*}
\subsection{Assumptions for Spatiotemporal Super-Resolution}
The proposed method is subject to specific assumptions and constraints. The deterministic generative event model~\cite{zhang2024tip}
\begin{equation}
    \Delta \mathbf{I}(t,\mathbf{x}) \approx - \left\langle\nabla_{\mathbf{x}}\mathbf{I}(t,\mathbf{x}), \mathbf{u}\Delta t\right\rangle
\end{equation}
indicates that events are triggered at the edge of an imaging object that is moving over a distance $\Delta \mathbf{x}=\mathbf{u}\Delta t$, where 
\begin{equation}
    \mathbf{I}(t,\mathbf{x}) = \int_{0}^{t} \tilde{\mathbf{E}}(\tau, \mathbf{x})~\mathrm{d}\tau
\end{equation}
is the quantized logarithmic intensity, and $\mathbf{u}$ denotes the motion field. With that, we assume:
\begin{enumerate}
    \item The scope is confined to the 2D projection of a 3D scene flow, where vertical motion along the $z$-axis is neglected.
    
    \item $\mathbf{u}$ is invariant within a short period $\Delta t$ such that there is a uniform event distribution in a small distance $\Delta \mathbf{x}$ (\eg, subpixels between $\mathbf{x}_0$ and $\mathbf{x}_1$). 

    \item $c$ is global and constant, which makes~\cref{eq:lx} hold. Despite it practically fluctuating with illumination and across pixels~\cite{brandli2014iscas}, the error induced can be minimized through network optimization such that satisfactory results can still be obtained on real-world samples through much simpler computations.
\end{enumerate}

\section{Experiments}
\subsection{Neuromorphic Super-Resolution}
\subsubsection{Implementation Details and Criterion}
The prototype architecture consists of a shallow 8-layer 3D CNN and an 11-layer MLP. Adam serves as the optimizer for training the two structures~\cite{kingma2015iclr}, for both with the learning rate of $10^{-3}$ being decayed by $0.1$ as per loss or epochs. We consult open-source materials to reproduce competing methods and conduct all experiments on PyTorch on NVIDIA GeForce RTX 3090 GPUs. Similar to~\cite{li2021iccv}, the raw recordings of a dataset, which are downsampled to synthesize LR counterparts, are taken as HR ground truth (HR-GT) in quantitative evaluations. We also follow~\cite{weng2022eccv} to use the root mean squared error (RMSE) as the assessment metric.

\subsubsection{Real Scenarios Visualization}
\cref{fig:vis_sr} shows visual results on challenging samples captured by a DAVIS240 camera~\cite{wang2020cvpr} and a DAVIS346 camera~\cite{zhang2023tci,zhang2024tip}, where we perform SR at the scale of $2 \times$ (from $190 \times 180$ to $380 \times 360$ pixels), $4 \times$ (from $105 \times 260$ to $420 \times 1040$ pixels), $8 \times$ (from $190 \times 180$ to $1520 \times 1440$ pixels), $16 \times$ (from $100 \times 70$ to $1600 \times 1120$ pixels), and $32 \times$ (from $130 \times 100$ to $4160 \times 3200$ pixels). The results from large-scale SR already exceed the resolution of the latest camera (\eg, Prophesee, EVK5, $1280 \times 720$ pixels). Our self-supervised mechanism is not subject to prior knowledge from external data and thus features an infinite SR function (in theory) that can reach a $32 \times$ scale, showing a higher degree of flexibility and practicability over the counterparts that fail (\ie, not applicable results) due to either high computational costs or training data unavailability. For full and zoom-in views at lower scales $2 \times$, $4 \times$, $8 \times$, we achieve more convincing results than the model-based one~\cite{li2019neurocomputing} and an equivalent reconstruction quality as~\cite{duan2021cvpr,weng2022eccv} that have seen a large quantity of instances in learning. More importantly, the generated subpixel events can enrich the visual texture/edge details that are insufficient in the LR records due to limited resolution or photon starvation in harsh illumination.

\begin{table}[t]
\caption{Quantitative evaluations on $2\times$ and $4\times$ neuromorphic SR.}
\label{table:ex_rmse_sr}
\centering
\begin{tabular}{llcccc}
    \toprule
    \textbf{Scale} & \textbf{Method} & \multicolumn{4}{c}{\textbf{RMSE}~$\downarrow$} \\ \midrule
     & & \texttt{poster} & \texttt{running} & \texttt{toy} & \texttt{text}\\ \cmidrule(lr){3-6}
    \multirow{5}{*}{$2 \times$} & Li~\etal~\cite{li2019neurocomputing} & $0.643$ & $0.488$ & $0.704$ & $0.427$\\
    &Duan~\etal~\cite{duan2021cvpr} & $\mathbf{0.547}$ & $0.363$ & $\mathbf{0.572}$ & $0.352$\\
    &Weng~\etal~\cite{weng2022eccv} & $0.581$ & $\mathbf{0.359}$ & $0.578$ & $0.326$\\\cmidrule(lr){3-6}
    &Ours& $0.569$ & $0.385$ & $0.592$ & $0.318$\\
    &Ours$^+$ & $0.552$ & $0.370$ & $0.581$ & $\mathbf{0.311}$\\ \midrule\midrule
    \multirow{5}{*}{$4 \times$} & Li~\etal~\cite{li2019neurocomputing} & $0.712$ & $0.539$ & $0.813$ & $0.445$\\
    &Duan~\etal~\cite{duan2021cvpr} & $0.605$ & $0.423$ & $0.626$ & $\mathbf{0.322}$\\
    &Weng~\etal~\cite{weng2022eccv} & $0.626$ & $\mathbf{0.417}$ & $0.633$ & $0.341$\\\cmidrule(lr){3-6}
    &Ours& $0.593$ & $0.462$ & $0.637$ & $0.332$\\
    &Ours$^+$& $\mathbf{0.590}$ & $0.439$ &$\mathbf{0.620}$ & $0.325$\\ 
    \bottomrule
\end{tabular}
\end{table}
\begin{table}[t]
\caption{$4\times$ neuromorphic SR for different downsampling methods.}
\label{table:downsample}
\centering
\begin{tabular}{llccc}
    \toprule
    \textbf{Downsample} & \textbf{Method} & \multicolumn{3}{c}{\textbf{RMSE}~$\downarrow$} \\ \midrule
     & & \texttt{poster}  & \texttt{toy} & \texttt{text}\\ \cmidrule(lr){3-5}
    \multirow{2}{*}{\texttt{Bicubic}} & Duan~\etal~\cite{duan2021cvpr} & $0.605$  & $\mathbf{0.626}$ & $\mathbf{0.322}$\\
    &Ours& $\mathbf{0.593}$  & $0.637$ & $0.332$\\\midrule\midrule
    \multirow{2}{*}{\texttt{Bilinear}} &Duan~\etal~\cite{duan2021cvpr} & $0.632$  & $0.683$ & $0.357$\\
    &Ours& $\mathbf{0.587}$  & $\mathbf{0.629}$ & $\mathbf{0.340}$\\\midrule\midrule
    \multirow{2}{*}{\texttt{Random}} &Duan~\etal~\cite{duan2021cvpr} & $0.641$  & $0.659$ & $0.368$\\
    &Ours& $\mathbf{0.607}$  & $\mathbf{0.650}$ & $\mathbf{0.349}$\\\bottomrule
\end{tabular}
\end{table}
\Cref{table:ex_rmse_sr} has quantitative analysis on the real-world recordings \texttt{poster\_6dof} and \texttt{outdoors\_running}~\cite{mueggler2017ijrr}, \texttt{toy} and \texttt{text\_intro}~\cite{zhang2024tip}. Besides, an ablation study simply upgrades our prototype by an advanced 3D U-Net structure~\cite{duan2021cvpr} and a deeper 20-layer MLP, denoted as Ours$^+$, for showing its potential and extensibility. Numerical comparisons demonstrate that ours promotes flexibility without significantly compromising accuracy. In addition, we investigate the impact of different downsampling methods on task performance. A benefit of self-supervised mechanisms over supervised ones is the stronger adaptability to various types of degradation and conditions. This is especially true for neuromorphic SR due to the dynamic acquisition by neuromorphic imaging. In contrast to supervised fashions requiring optimization on a fixed setting, ours can flexibly adapt to the specific degradation of a test sample, at test time. \Cref{table:downsample} assesses one ideal case (bicubic) and two non-ideal cases (bilinear, random). These known kernels do not significantly affect our performance since the model already has such knowledge in inference (\ie, \cref{eq:downsample}), while the supervised counterpart trained on the bicubic kernel underperforms for other degradation scenarios. 

\begin{figure}[t]
    \centering
    \subfloat[LR source]{\includegraphics[width=110pt, height=90pt]{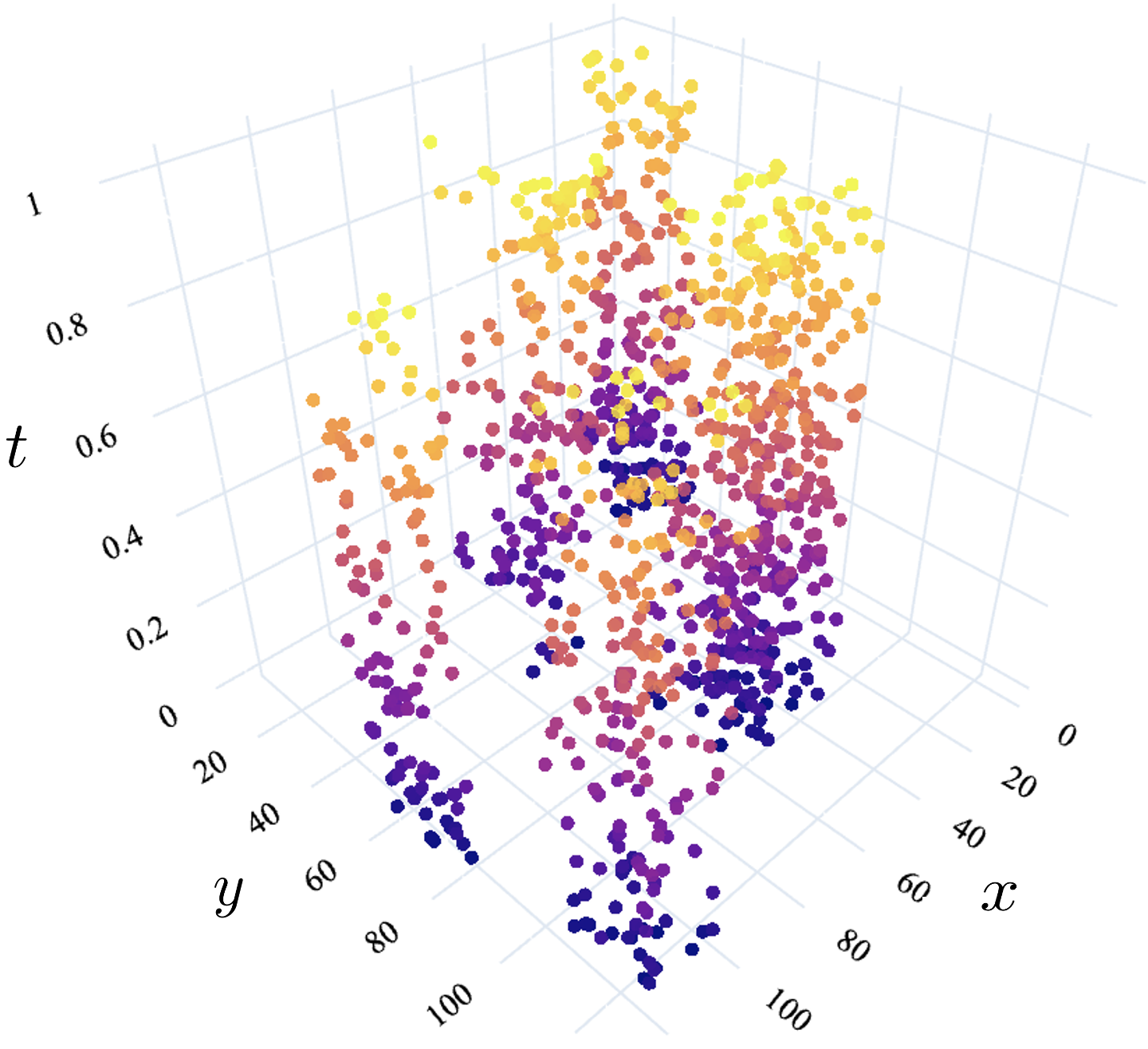}}\hspace{4pt}
    \subfloat[$2\times$ SR]{\includegraphics[width=110pt, height=90pt]{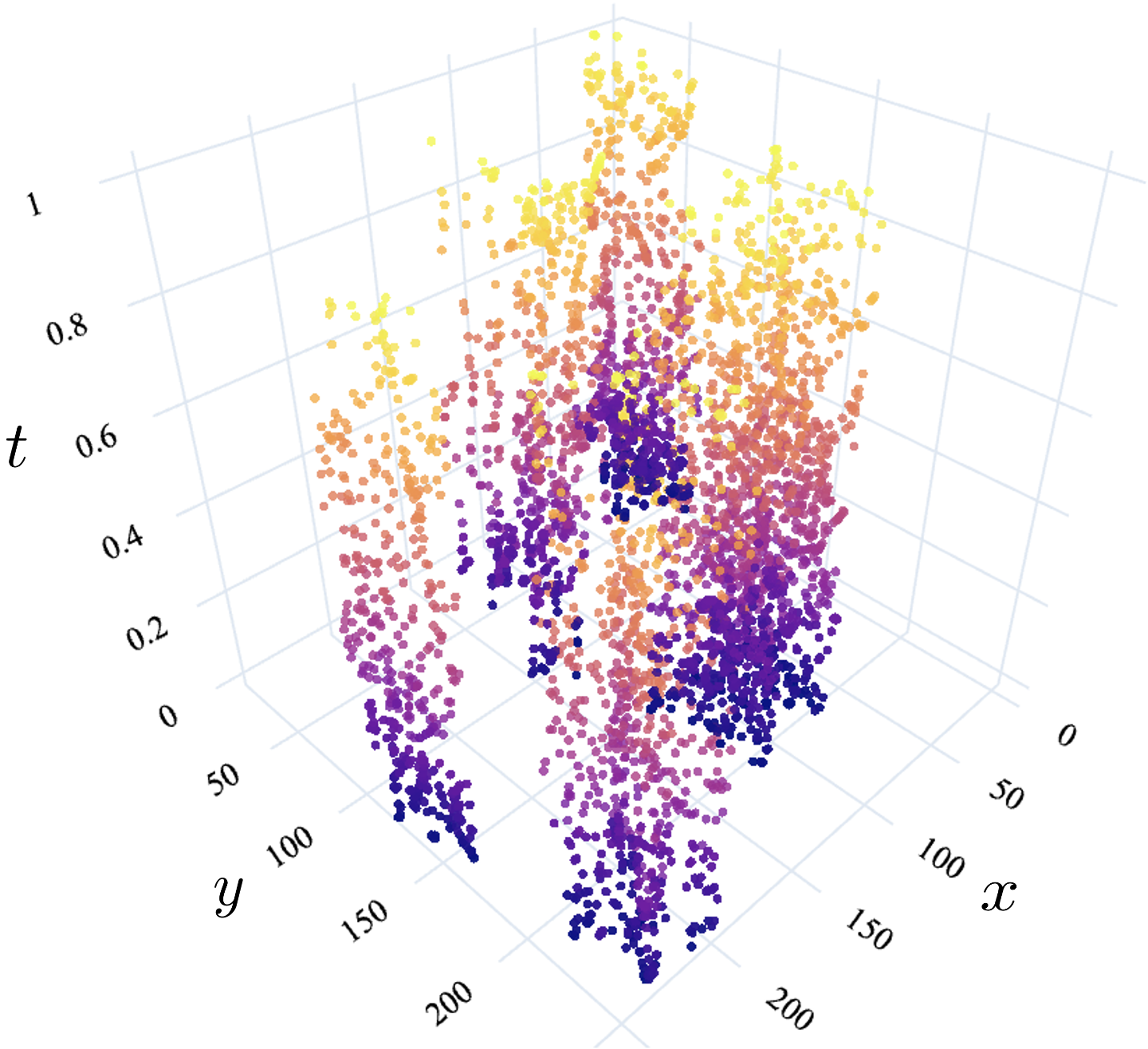}}\\
    \subfloat[Spatiotemporal distribution]{\includegraphics[width=250pt]{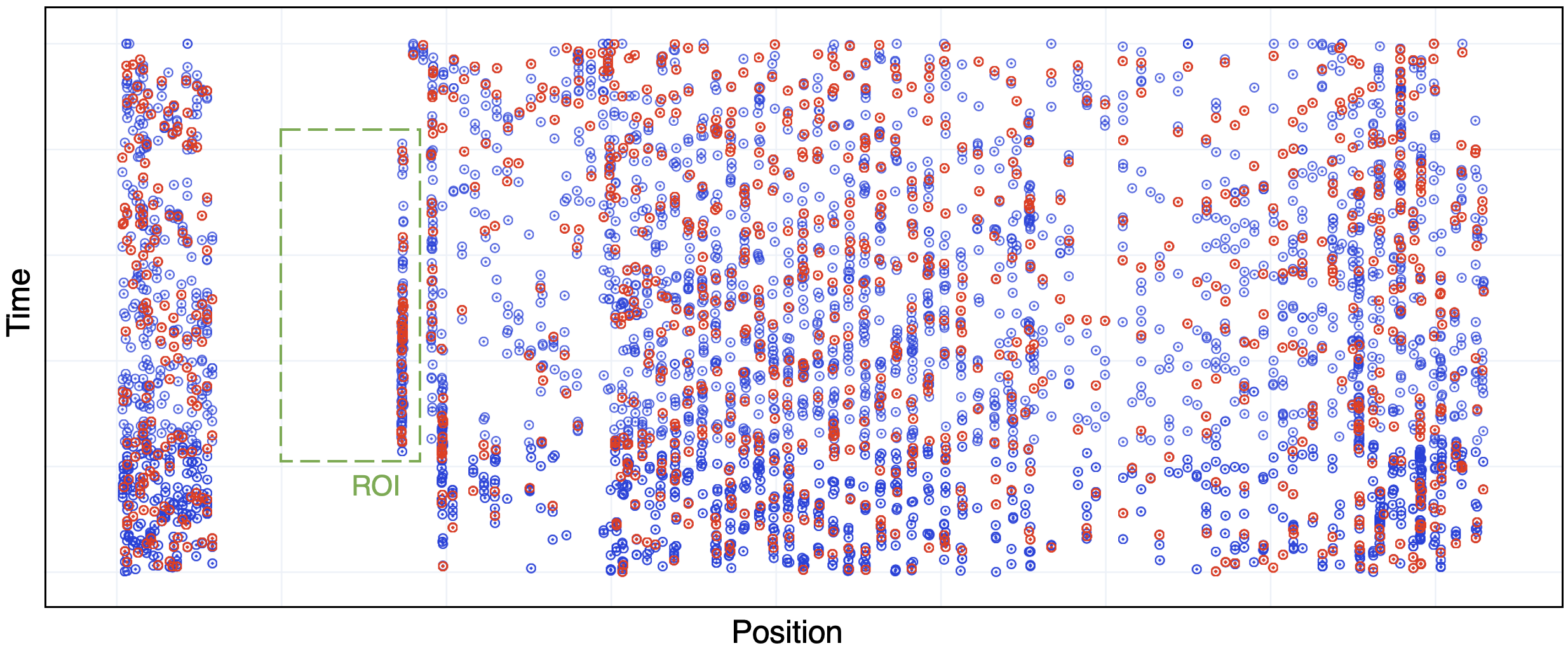}}\hfill
    \caption{(a)--(b) 3D visualization of LR events and our SR result (brighter ones have more recent time). (c) The corresponding spatiotemporal distribution, where red dots for (a) and blue dots for (b).}
    \label{fig:vis_sr3d}
\end{figure}
Flattening events into a 2D plane hardly reveals temporal variations that are key to distinguish between frame- and event-level SR. \cref{fig:vis_sr3d} (a) and (b) visualize a LR sample~\cite{orchard2015tpami} and our SR estimate in a 3D view. Apart from maintaining similar characteristics in space-time, our method is also found to function temporal upsampling~\cite{xiang2022icme} and sparse event completion~\cite{zhang2024sr}, where events become much denser along both spatial and temporal axes, augmenting the quantity of highly sparse events from a weakly dynamic scene. In addition, the corresponding distribution is plotted in~\cref{fig:vis_sr3d} (c). The green-marked ROI shows events triggered at different time at a close position to highlight the precision of our spatial SR, where the generated events do not have a significant spatial-offset from the LR raw. The time density of both also follows a similar pattern --- more frequent LR events generally lead to more after temporal SR.

\begin{figure}[t]
    \centering
    \subfloat[Recognition on $2\times$ SR events of \texttt{ASL-DVS}~\cite{bi2019iccv} and \texttt{N-CARS}~\cite{sironi2018cvpr}]{\includegraphics[width=0.4\textwidth]{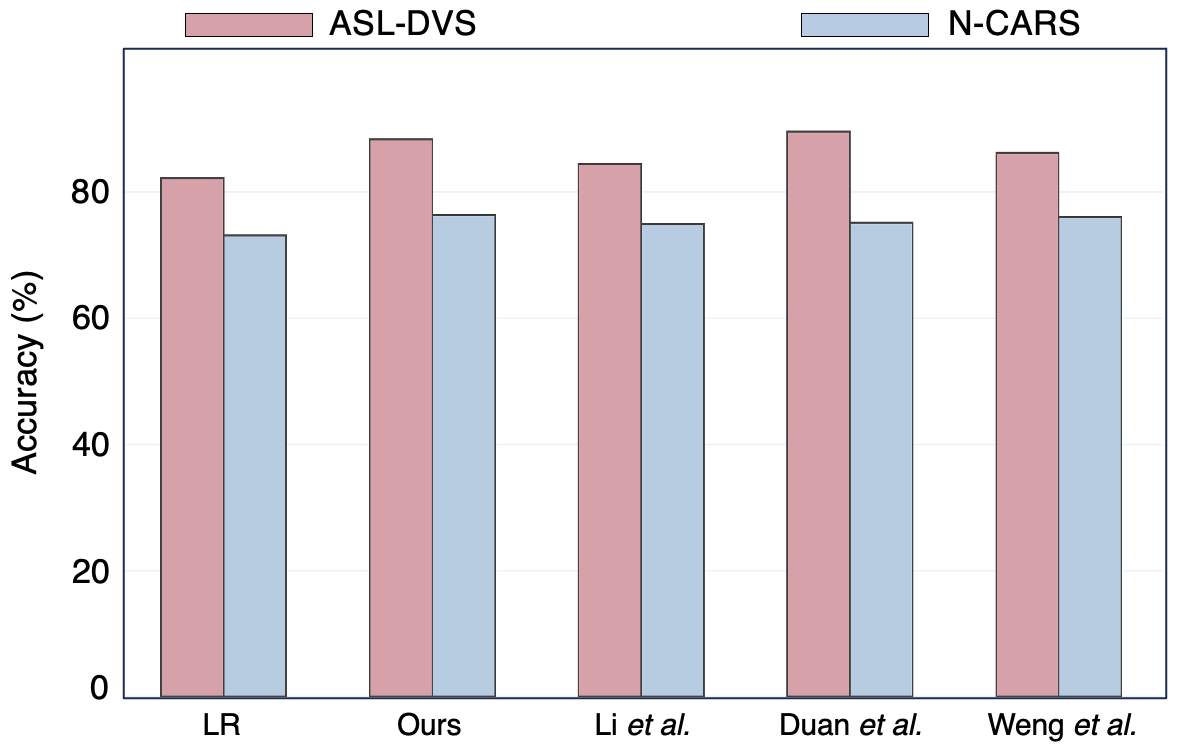}}\\
    \subfloat[Detection on $2\times$, $4\times$ SR events of \texttt{GEN1}~\cite{de2020large}]{\includegraphics[width=0.4\textwidth]{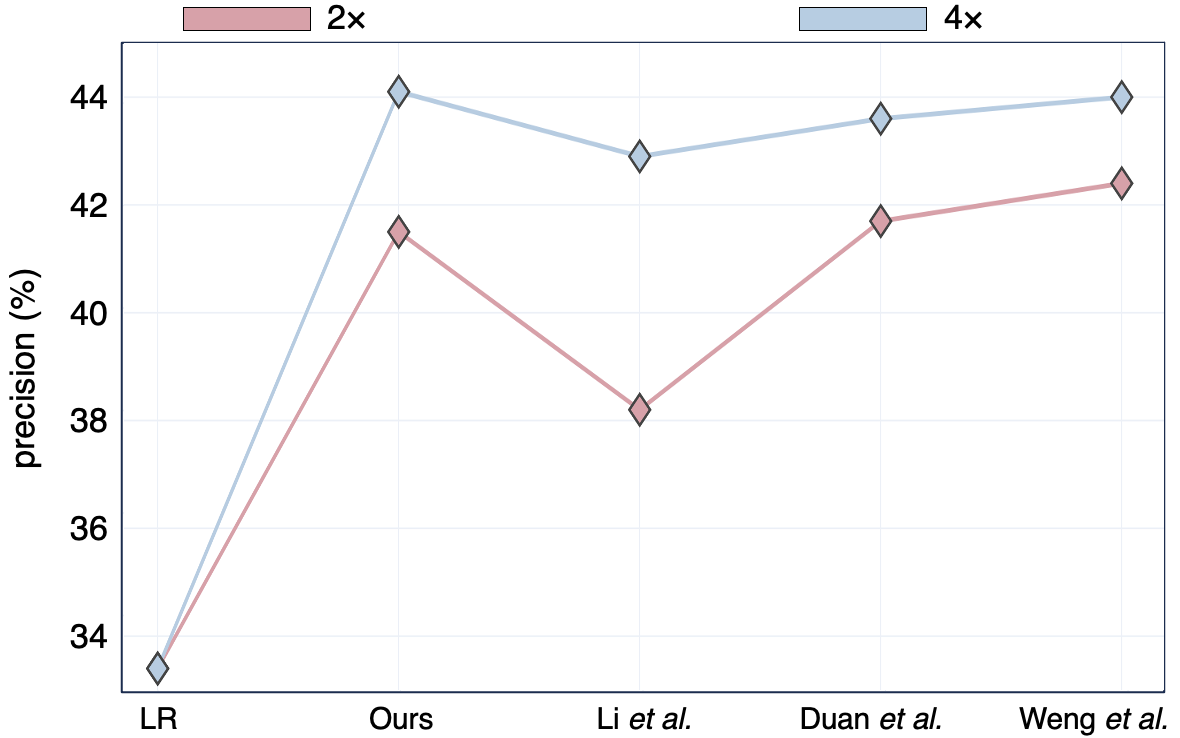}}
    \caption{Neuromorphic SR for object recognition and detection.}
    \label{fig:sr_reasoning}
\end{figure}
\subsubsection{Neuromorphic Reasoning}
High-level neuromorphic reasoning on an event stream is an effective way to reveal its underlying patterns that cannot be observed through visualization. First, we investigate whether $2 \times$ SR has positive impacts on recognition by evaluating a benchmark classifier~\cite{gehrig2019iccv} on the streams of \texttt{ASL-DVS}~\cite{bi2019iccv} and \texttt{N-CARS}~\cite{sironi2018cvpr}. In~\cref{fig:sr_reasoning}~(a), the average accuracy increases when feeding SR samples. Compared with the first dataset collected in a laboratory setup, fewer gains are obtained for \texttt{N-CARS} with a higher noise level. One reason might be the dramatically grown noise quantity in SR estimates. Our self-supervised method, without being trapped by any prior, holds a competitive edge in this particular case. In~\cref{fig:sr_reasoning}~(b), we analyze object detection tasks~\cite{lin2014eccv}, where a backbone~\cite{gehrig2023cvpr} evaluates LR and $2 \times$, $4 \times$ SR events on \texttt{GEN1} ($304 \times 240$ pixels)~\cite{de2020large}. Surprisingly, the precision grow is significant for $2 \times$ SR yet marginal in $4 \times$ SR cases. There might be an upper resolution bound in which an algorithm maximizes its performance. $2 \times$ SR samples have a proper spatial resolution sufficient for high-quality reasoning, and further augmentation hardly pushes one to extract more features from events. This observation also raises a rethinking of the optimal camera resolution for various use cases, due to trade-offs between computing resources and desired precision. Above studies demonstrate that neuromorphic SR can elevate downstream tasks to a certain extent, and our method achieves highly competitive results compared with the state-of-the-arts.

\begin{table}[t]
\caption{Runtime (in seconds) analysis and comparisons.}
\label{table:runtime}
\centering
\begin{tabular}{llllr}
    \toprule
    \textbf{Scale} & \textbf{Time} & \multicolumn{3}{c}{\textbf{Method}} \\ \midrule
     & & Duan~\etal~\cite{duan2021cvpr} & Weng~\etal~\cite{weng2022eccv} & Ours\\ \cmidrule(lr){3-5}
    \multirow{2}{*}{$2 \times (34)$} & Train & \textbackslash & \textbackslash &  $24.1+2.8$\\
    &Test & $0.14$ & $0.10$ &  $0.2+0.01$\\\midrule\midrule
    \multirow{2}{*}{$2 \times (128)$} & Train & \textbackslash & \textbackslash &  $30.3+43.5$\\
    &Test & $0.27$ & $0.19$ & $1.1+0.02$\\\midrule\midrule
    \multirow{2}{*}{$4 \times (128)$} & Train & \textbackslash & \textbackslash &  $34.2+44.8$\\
    &Test & $0.36$ & $0.25$ &  $2.8+0.06$\\\midrule\midrule
    \multirow{2}{*}{$4 \times (346)$} & Train & \textbackslash & \textbackslash & $26.8+62.7$ \\
    &Test & $0.54$ & $0.40$ &  $7.2+0.13$\\
    \bottomrule
\end{tabular}
\end{table}
\subsubsection{Runtime}
Although our self-supervised method is trained at test time, its overall runtime, which influences the practical applicability in real-world scenarios, deserves clarification and discussion. \Cref{table:runtime} investigates the runtime required for the samples with different spatial resolutions ($2\times$ SR for $34\times34$ and $128\times128$ pixels, $4\times$ SR for $128\times128$ and $346\times260$ pixels). Our training time is allocated for spatial SR ($1000$ iterations) and temporal SR ($1000$ epochs). The former, which fluctuates around $25$--$35$\si{\s}, is almost independent of the input resolution and the scaling factor, whereas the latter only grows as the resolution increases. For example, for the same input with $128\times128$ pixels, temporal SR takes similar training time \SI{43.5}{\s} for $2\times$ and \SI{44.8}{\s} for $4\times$. In contrast, the sample with $34\times34$ pixels has mere \SI{2.8}{\s}, and the one with $346\times260$ pixels consumes \SI{62.7}{\s}. The higher the resolution, the more mappings to be learnt (\ie, \cref{eq:loss_t}). As for inference time, both rise with the resolution and the scaling factor, but the latter is negligible. In addition, we measure the supervised counterparts for comparisons. As expected, ours is at a disadvantage in terms of spatial SR. The depth of our event voxel-grid depends on the largest size of an event stream (\ie, \cref{eq:lx}), which is much deeper than the counterparts that have a reduced depth dimension with feature loss. It thus demands more time as a result of more computations. Nevertheless, our approach still has acceptable latency and a trivial impact on most scenarios.

\begin{figure}[t]
    \centering
    \subfloat[Raw image]{\includegraphics[width=80pt]{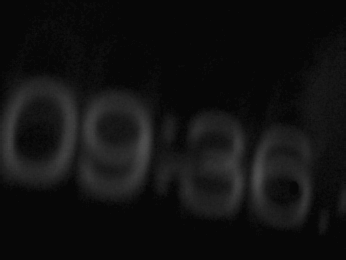}}\hfill
    \subfloat[$2\times$ SR]{\includegraphics[width=80pt]{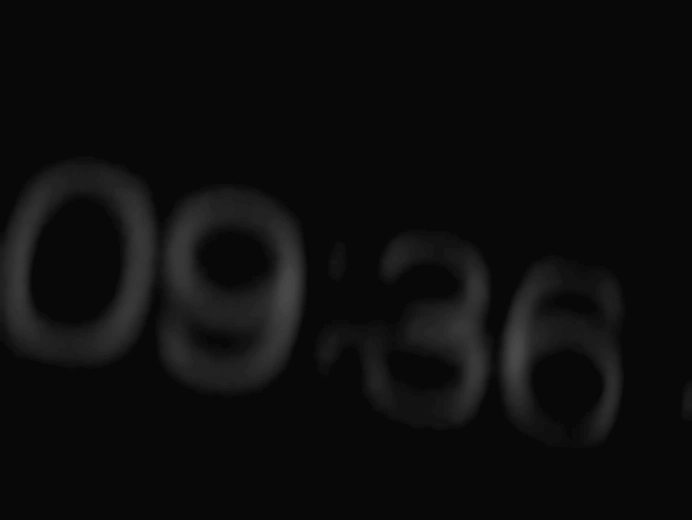}}\hfill
    \subfloat[$4\times$ SR]{\includegraphics[width=80pt]{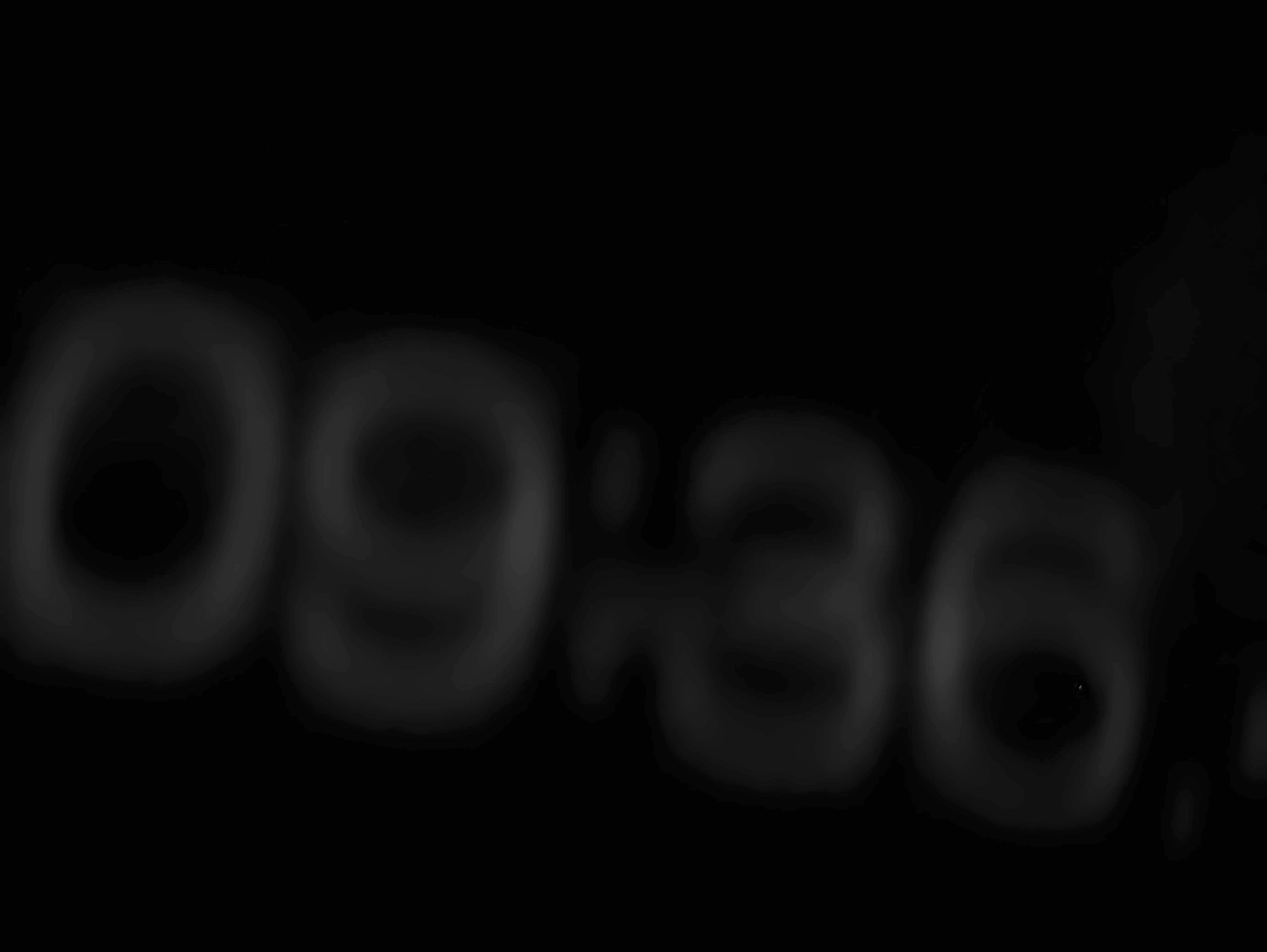}}
    \caption{Image SR methods underperform for underexposed, blurry images.}
    \label{fig:img_sr}
\end{figure}
\begin{figure*}[t]
    \subfloat[Neuromorphic imaging setup]{\includegraphics[width=215pt, height=268pt]{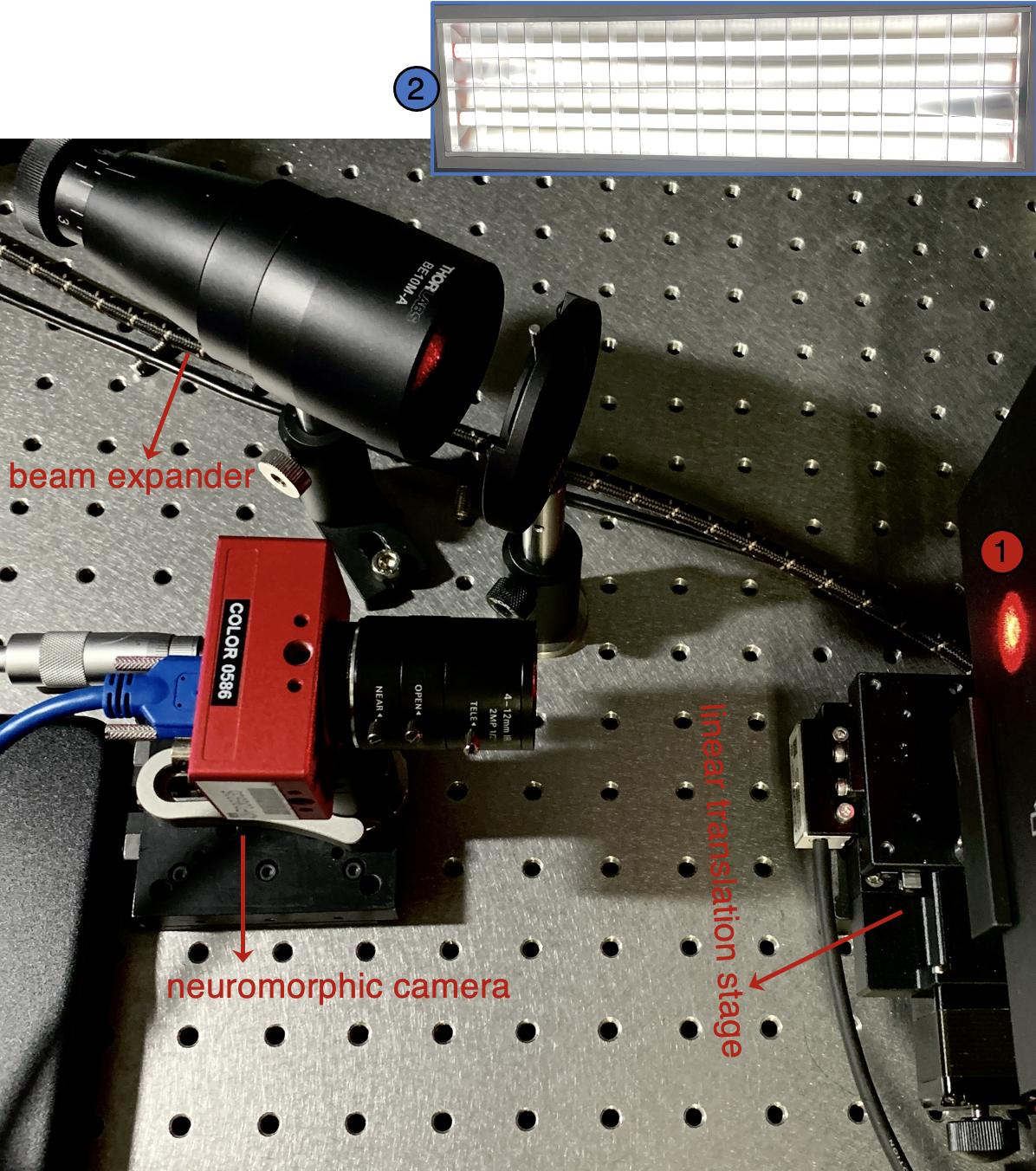}}\hspace{-8pt}
    \subfloat[LR sources]{
    \begin{tabular}[b]{c}
        \frame{\includegraphics[width=66pt, height=50pt]{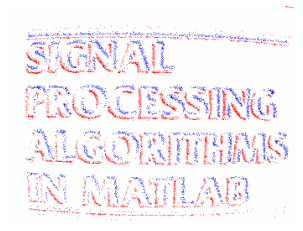}}\\
        \frame{\includegraphics[width=66pt, height=50pt]{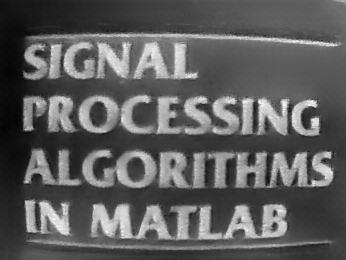}}\\
        \begin{tabular}[b]{c}
        \frame{\includegraphics[width=31pt, height=25pt]{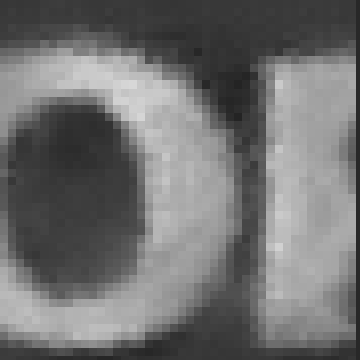}}
        \frame{\includegraphics[width=31pt, height=25pt]{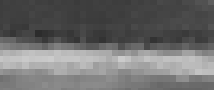}}      
        \end{tabular}\\
        \frame{\includegraphics[width=66pt, height=50pt]{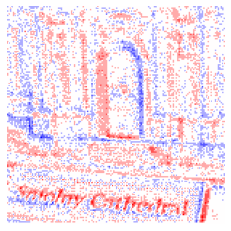}}\\
        \frame{\includegraphics[width=66pt, height=50pt]{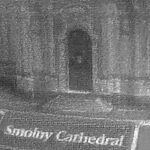}}\\
        \begin{tabular}[b]{c}
        \frame{\includegraphics[width=31pt, height=25pt]{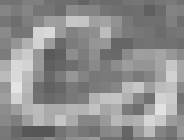}}
        \frame{\includegraphics[width=31pt, height=25pt]{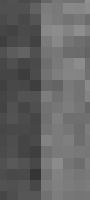}}      
        \end{tabular}
    \end{tabular}}\hspace{-22pt}%
    \subfloat[Ours]{
    \begin{tabular}[b]{c}%
        \frame{\includegraphics[width=66pt, height=50pt]{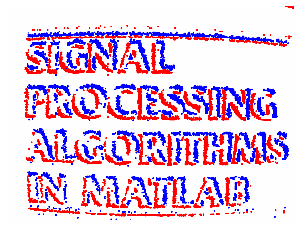}}\\
        \frame{\includegraphics[width=66pt, height=50pt]{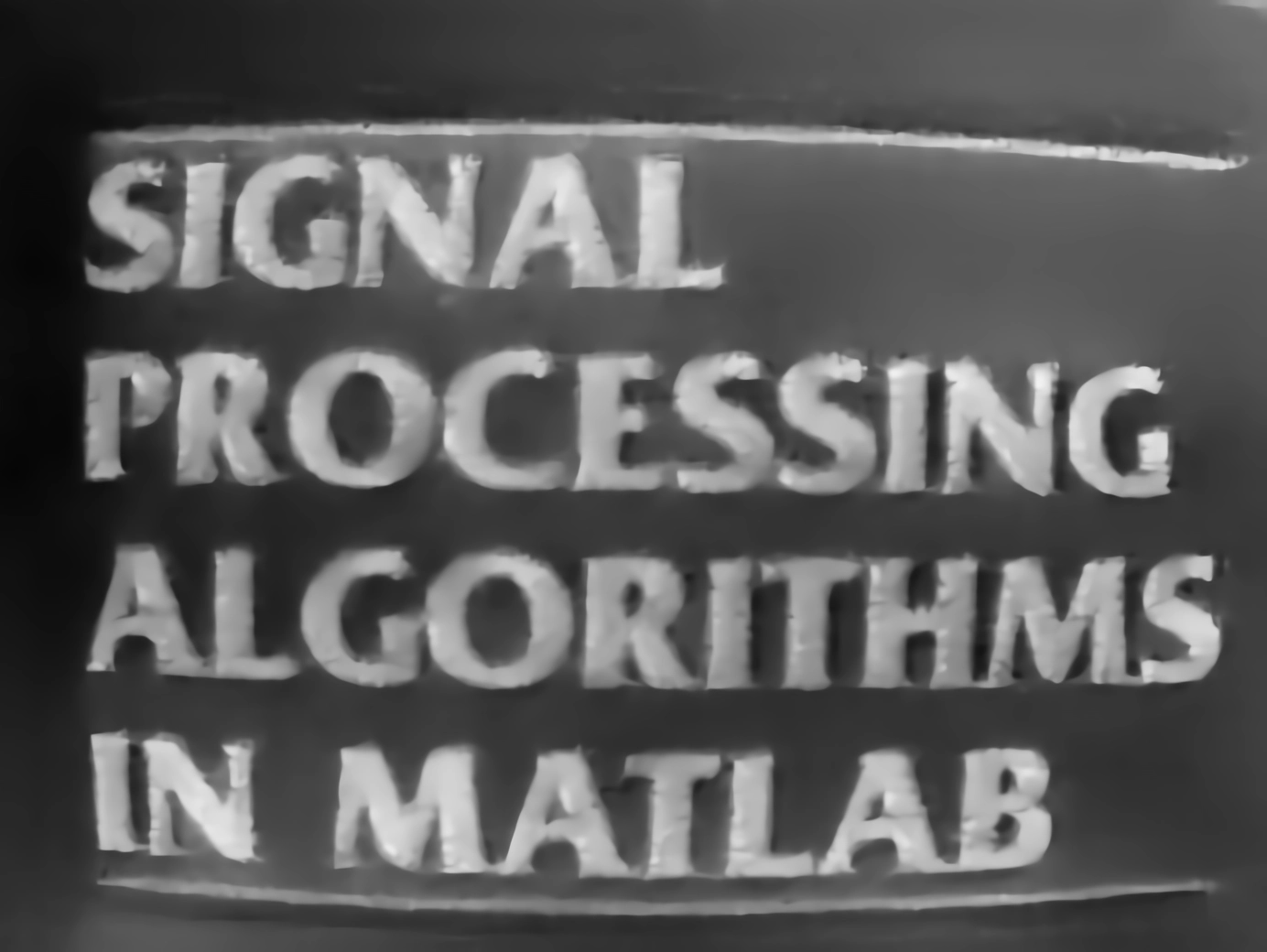}}\\
        \begin{tabular}[b]{c}
        \frame{\includegraphics[width=31pt, height=25pt]{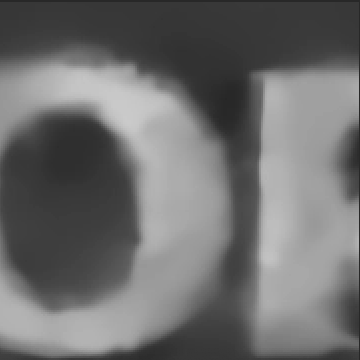}}
        \frame{\includegraphics[width=31pt, height=25pt]{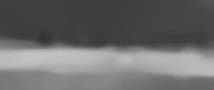}}      
        \end{tabular}\\
        \frame{\includegraphics[width=66pt, height=50pt]{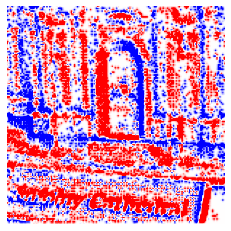}}\\
        \frame{\includegraphics[width=66pt, height=50pt]{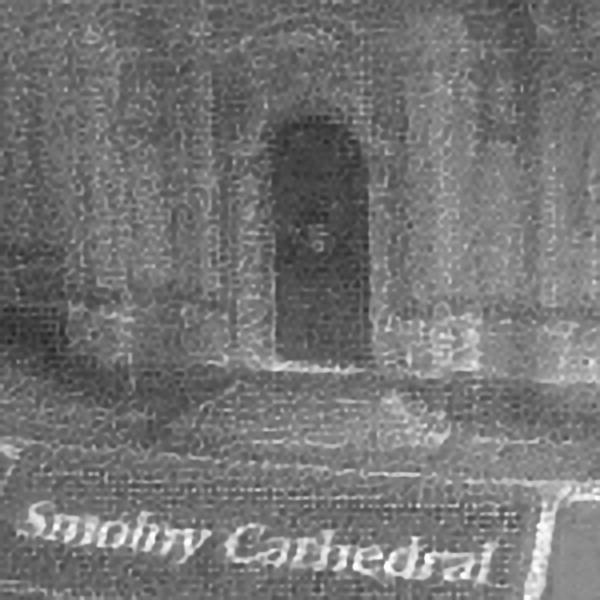}}\\
        \begin{tabular}[b]{c}
        \frame{\includegraphics[width=31pt, height=25pt]{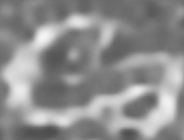}}
        \frame{\includegraphics[width=31pt, height=25pt]{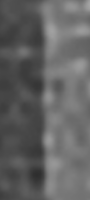}}      
        \end{tabular}
    \end{tabular}}\hspace{-22pt}%
    \subfloat[Li~\etal~\cite{li2019neurocomputing}]{
    \begin{tabular}[b]{c}%
        \frame{\includegraphics[width=66pt, height=50pt]{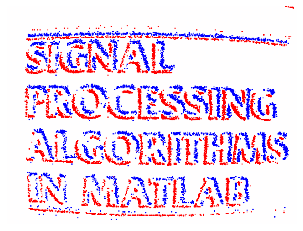}}\\
        \frame{\includegraphics[width=66pt, height=50pt]{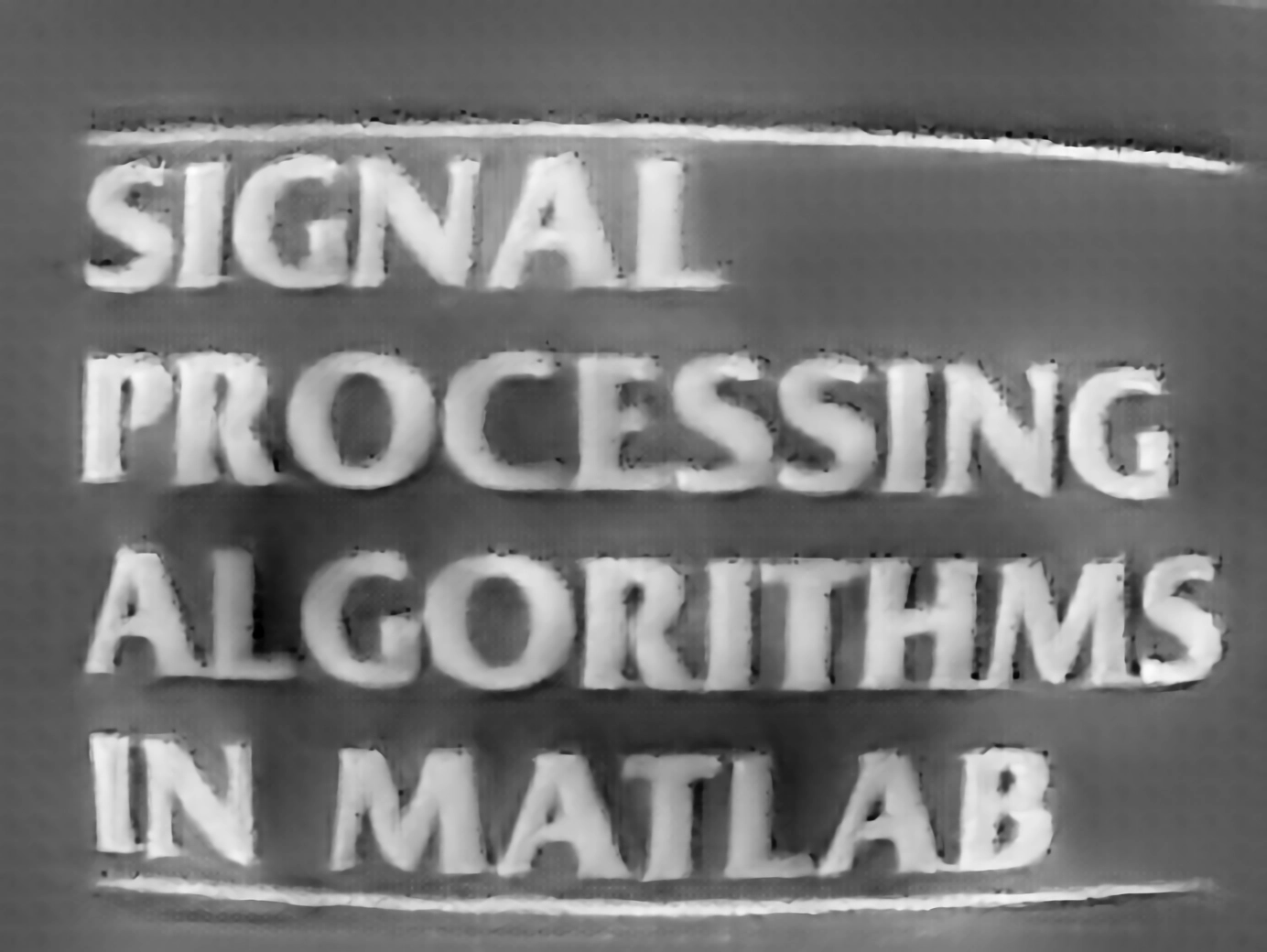}}\\
        \begin{tabular}[b]{c}
        \frame{\includegraphics[width=31pt, height=25pt]{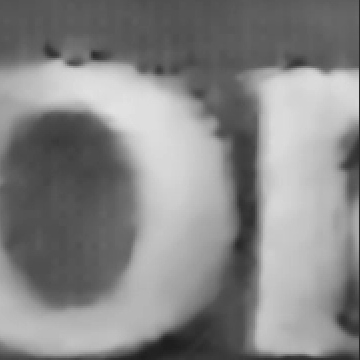}}
        \frame{\includegraphics[width=31pt, height=25pt]{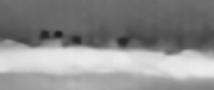}}      
        \end{tabular}\\
        \frame{\includegraphics[width=66pt, height=50pt]{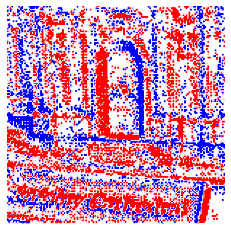}}\\
        \frame{\includegraphics[width=66pt, height=50pt]{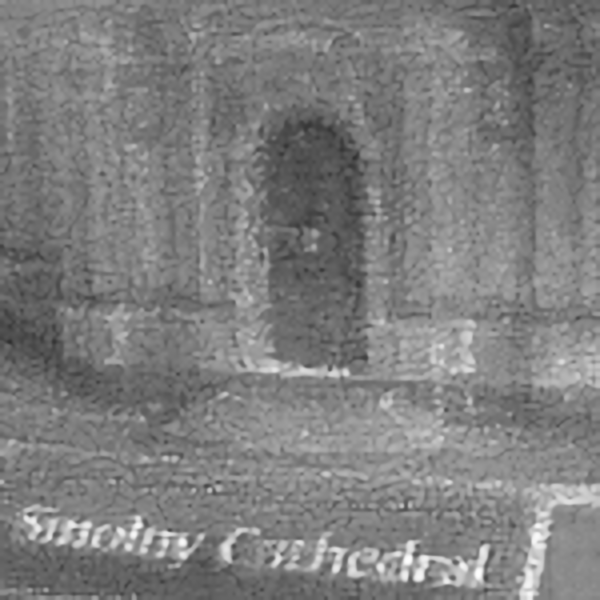}}\\
        \begin{tabular}[b]{c}
        \frame{\includegraphics[width=31pt, height=25pt]{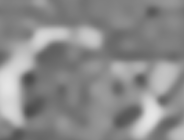}}
        \frame{\includegraphics[width=31pt, height=25pt]{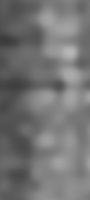}}      
        \end{tabular}
    \end{tabular}}\hspace{-22pt}%
    \subfloat[Weng~\etal~\cite{weng2022eccv}]{
    \begin{tabular}[b]{c}%
        \frame{\includegraphics[width=66pt, height=50pt]{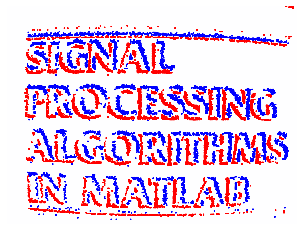}}\\
        \frame{\includegraphics[width=66pt, height=50pt]{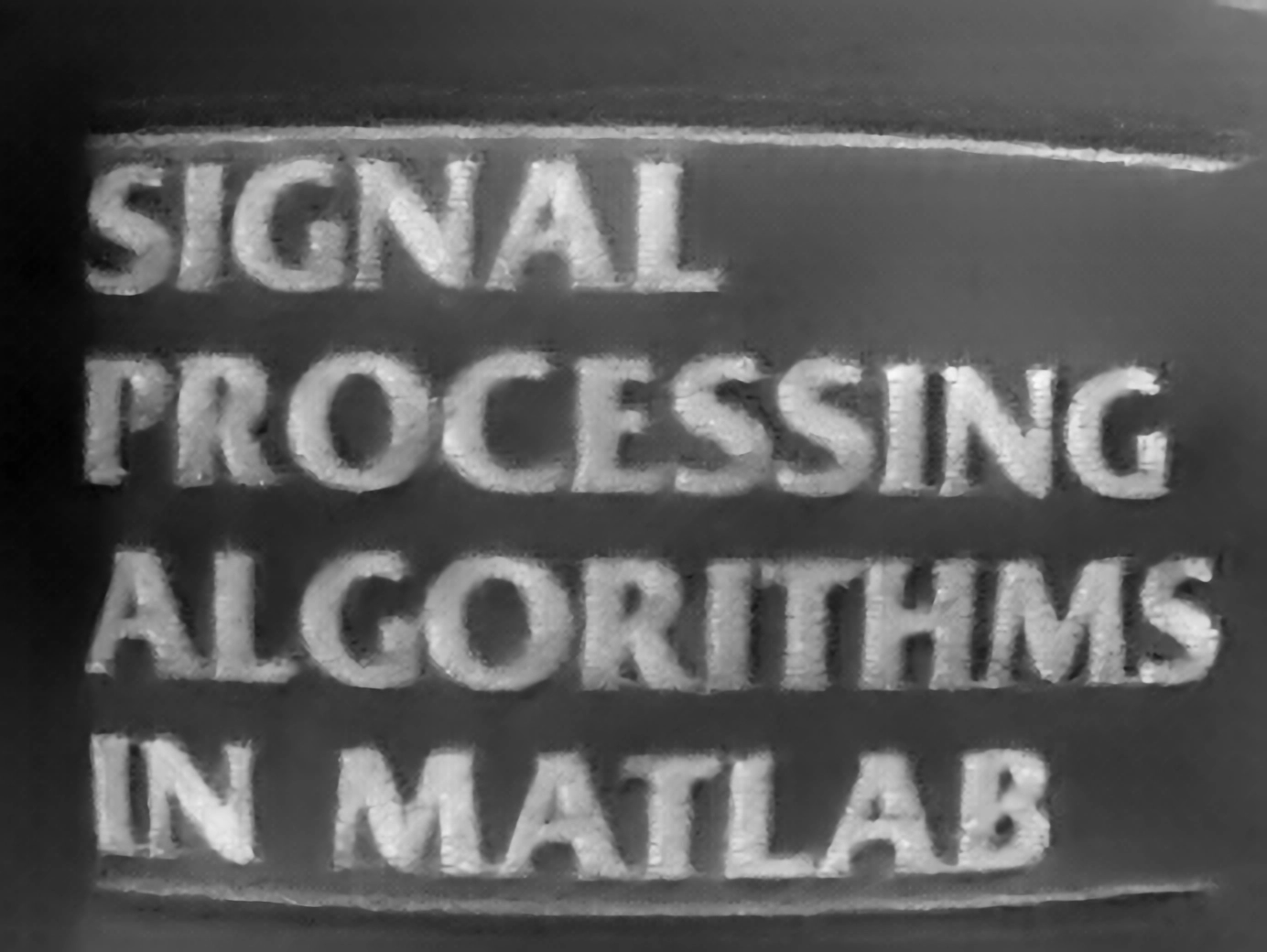}}\\
        \begin{tabular}[b]{c}
        \frame{\includegraphics[width=31pt, height=25pt]{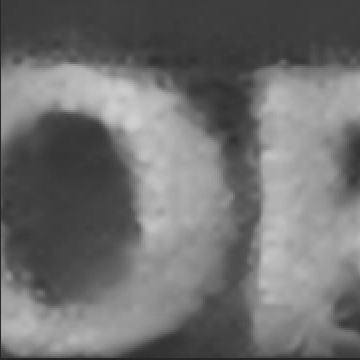}}
        \frame{\includegraphics[width=31pt, height=25pt]{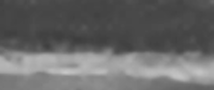}}      
        \end{tabular}\\
        \frame{\includegraphics[width=66pt, height=50pt]{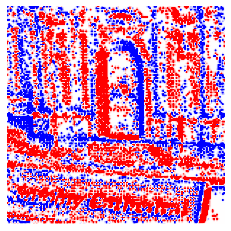}}\\
        \frame{\includegraphics[width=66pt, height=50pt]{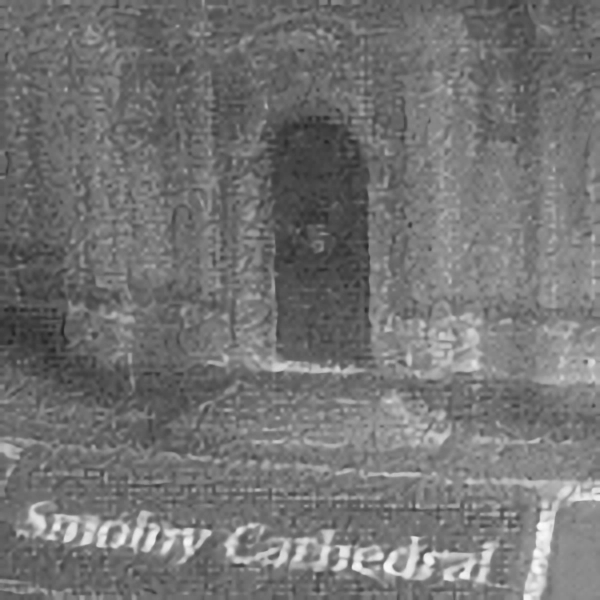}}\\
        \begin{tabular}[b]{c}
        \frame{\includegraphics[width=31pt, height=25pt]{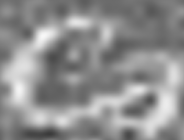}}
        \frame{\includegraphics[width=31pt, height=25pt]{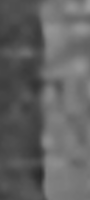}}      
        \end{tabular}
    \end{tabular}}
    \caption{(a) Two kinds of experimental setup for neuromorphic imaging. (b)--(e) LR ($346 \times 260$ pixels), $4\times$ SR ($1384 \times 1040$ pixels) events, and their reconstructed images accompanied by two focused ROIs.}
    \label{fig:img_rec}
\end{figure*}
\subsection{Improved Synergy with Frame Imaging}
One of the convincing motivations behind neuromorphic SR is the potential to achieve significantly improved synergy with frame imaging, in which SR events can unlock the capability for low-frequency signal reconstruction via upgraded clarity and sharpness. \textit{Why don't we use well-established image SR?} Frame cameras with a limited frame rate and a low dynamic range often capture inferior images of a blurry, underexposed or overexposed state, from which image SR fails to recover more information yet remains a bad quality at most. A sample under low-light imaging in~\cref{fig:img_sr} shows that pixels during the processing are gradually faded and distorted due to low contrast, posing a need for neuromorphic SR as an alternative.

\subsubsection{Natural Image}
\cref{fig:img_rec}~(a) presents two typical settings in a laboratory. The first, which yields almost noise-free events, comprises a HeNe laser (Thorlabs, HNL100L, $\lambda =\SI{632.8}{\nm}$) as a stable light source, a beam expander (Thorlabs, BE10M-A) evenly distributes the light across the region of motion, and a neuromorphic camera (iniVation, DAVIS346, $346 \times 260$ pixels) records a target mounted on a motorized linear translation stage (WN262TA20, Winner Optics). Another configuration that makes noisy events normally uses a fluorescent lamp as the lighting, which exhibits flickers of $100$ times per second due to the \SI{50}{\Hz} alternating current. Samples on a handheld rig have more irregular and complex movement.

\cref{fig:img_rec}~(b) compares a noise-free instance with a noisy one, and~\cref{fig:img_rec}~(c)--(e) visualize the $4\times$ SR (from $346 \times 260$ to $1384 \times 1040$ pixels), along with their reconstructed images. E2VID~\cite{rebecq2021pami} provides a dedicated event-to-image mapping. For both scenarios, the difference among the evaluated methods lies in event sparsity and continuity in certain ROIs, which is marginal and hard to observe in a frame-based form. However, incorporating temporal features, which associates with event correlation, to reconstruct images can magnify such variations including fidelity of structures, sharpness of edges, and shade of gray. Visual comparisons and zoom-in views present that our approach achieves quite satisfactory results. 

\begin{figure}[t]
    \centering
    \includegraphics[width=0.48\textwidth]{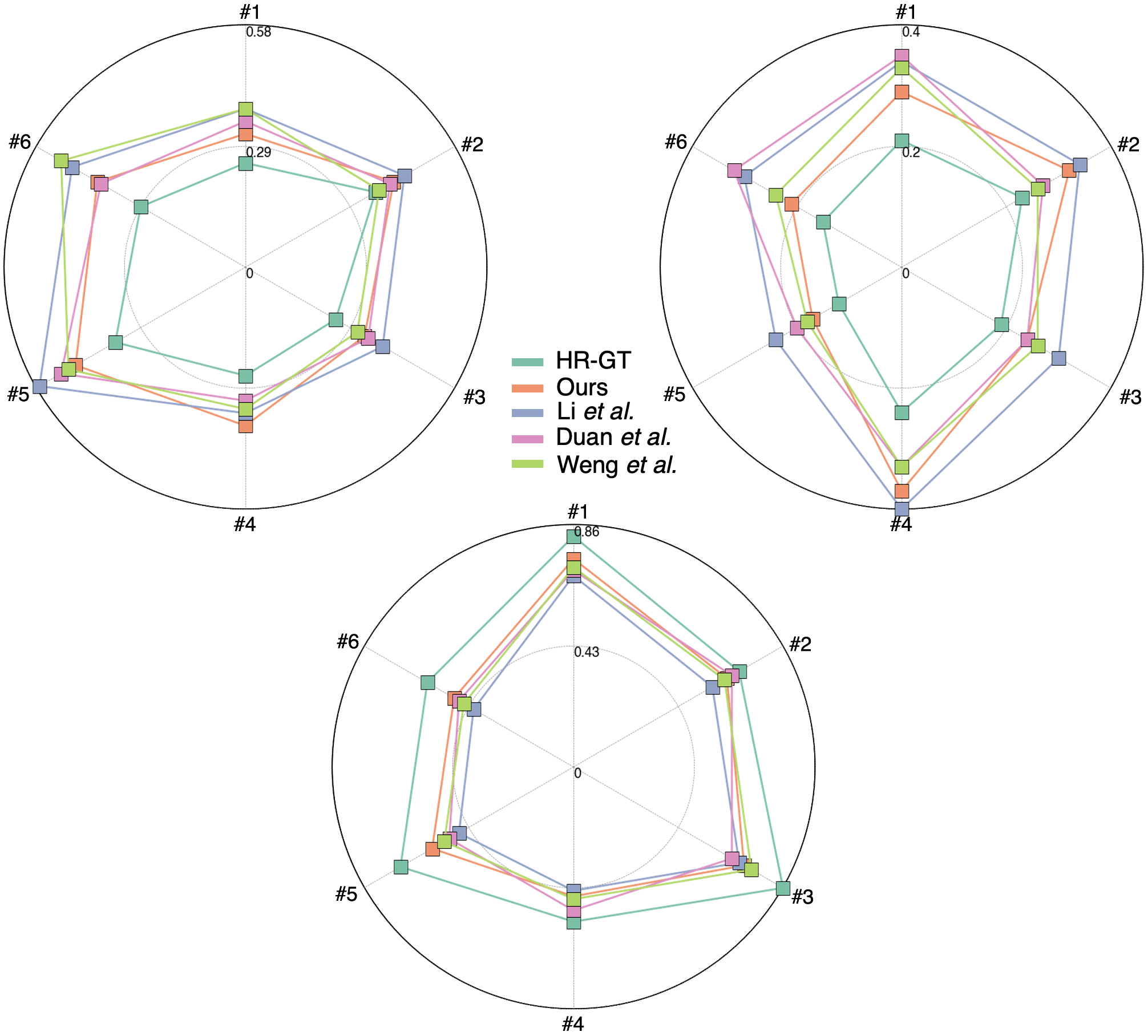}
    \caption{LPIPS~$\downarrow$ (left), MSE~$\downarrow$ (right), and SSIM~$\uparrow$ (bottom) assessments for reconstructed images of SR events.}
    \label{fig:img_rec_chart}
\end{figure}

\cref{fig:img_rec_chart} depicts LPIPS~\cite{zhang2018cvpr}, MSE, and SSIM evaluations on
\begin{enumerate*}
    \item \texttt{outdoors\_walking}
    \item \texttt{outdoors\_running}
    \item \texttt{shapes\_6dof}
    \item \texttt{dynamic\_6dof}
    \item \texttt{boxes\_6dof}
    \item \texttt{poster\_6dof}
\end{enumerate*}
\cite{mueggler2017ijrr}. Downsampled LR versions from raw recordings are upgraded to the corresponding SR estimates, whose reconstructed images are compared with those of the HR-GT. Quantitatively, ours is highly competitive with the state-of-the-arts on each measure in each sample.

\begin{figure*}[t]
    \centering    
    \subfloat[Neuromorphic microscopy setup]{\frame{\includegraphics[width=180pt, height=225pt]{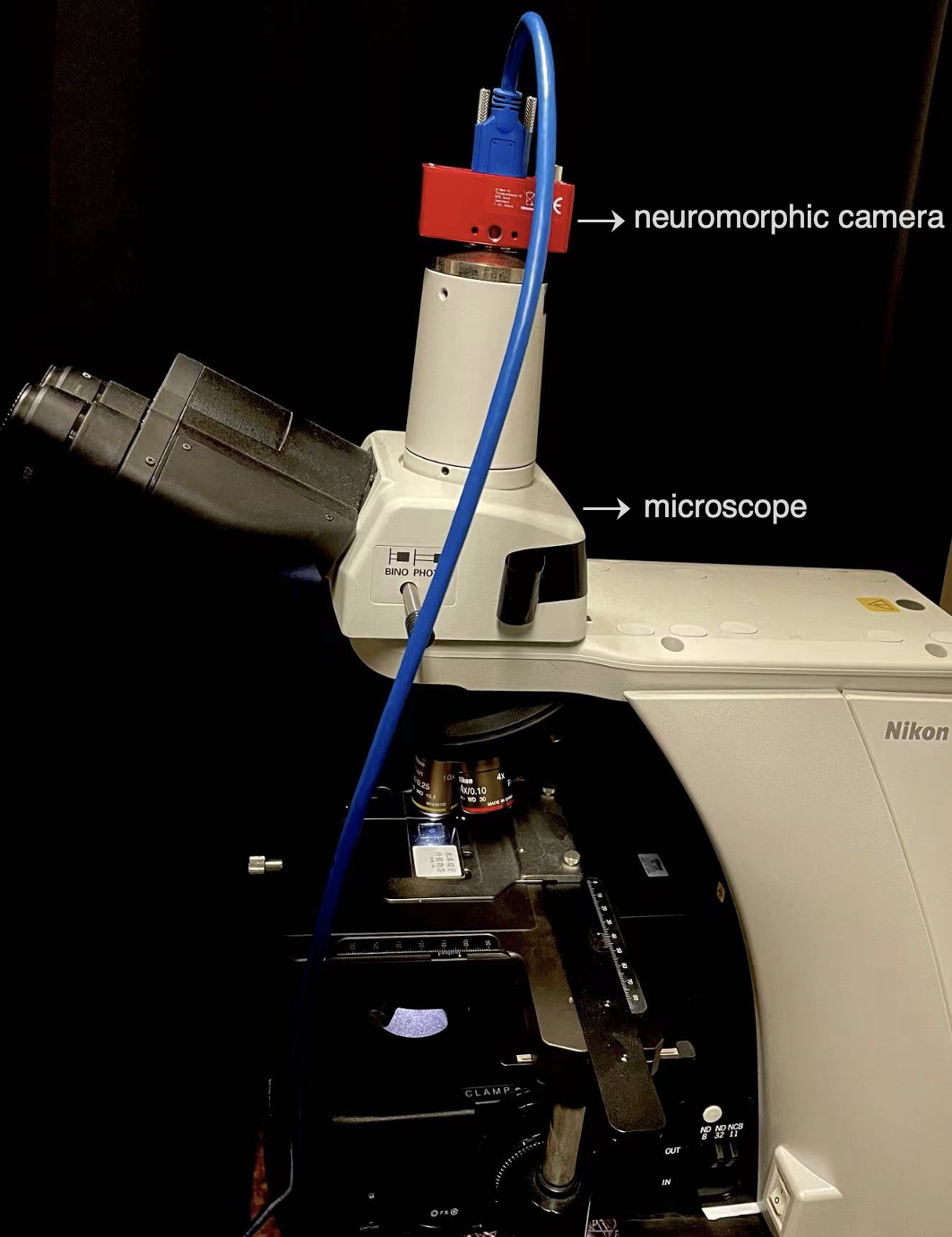}}}
    \begin{tabular}[b]{c}
    \subfloat[\texttt{Honeybee hindleg}, $10\times$ objective]{
    \frame{\includegraphics[width=76pt, height=60pt]{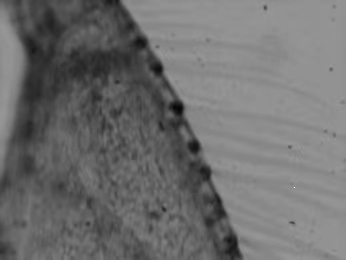}}
    \frame{\includegraphics[width=76pt, height=60pt]{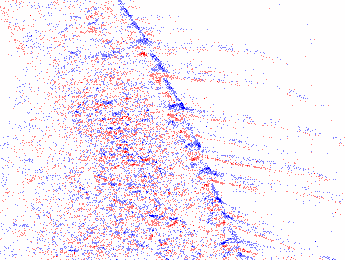}}
    \frame{\includegraphics[width=76pt, height=60pt]{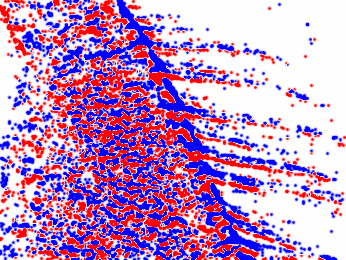}}
    \frame{\includegraphics[width=76pt, height=60pt]{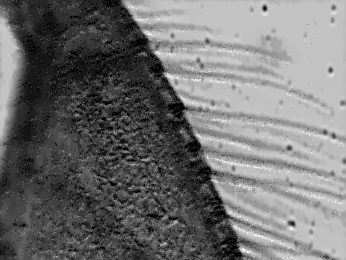}}
    }\\
    \subfloat[\texttt{Drosophila}, $2\times$ objective]{
    \stackinset{l}{0pt}{b}{0pt}{\frame{\includegraphics[width=22pt]{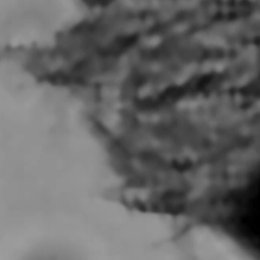}}}{\stackinset{l}{0pt}{t}{0pt}{\frame{\includegraphics[width=22pt]{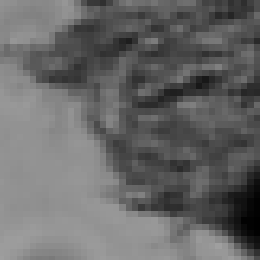}}}
    {\frame{\includegraphics[width=76pt, height=60pt]{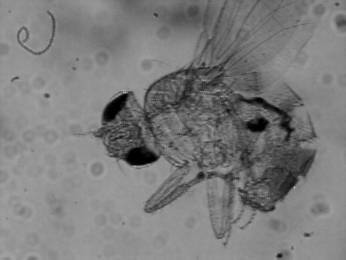}}}}
    \frame{\includegraphics[width=76pt, height=60pt]{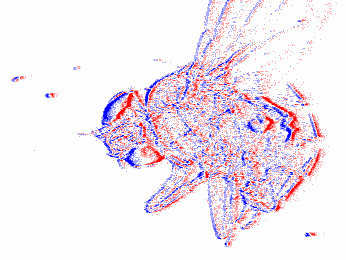}}
    \frame{\includegraphics[width=76pt, height=60pt]{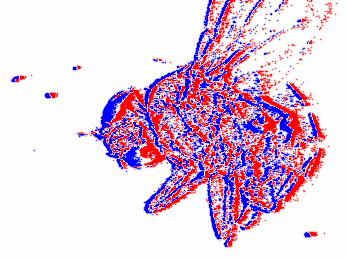}}
    \stackinset{l}{0pt}{b}{0pt}{\frame{\includegraphics[width=22pt]{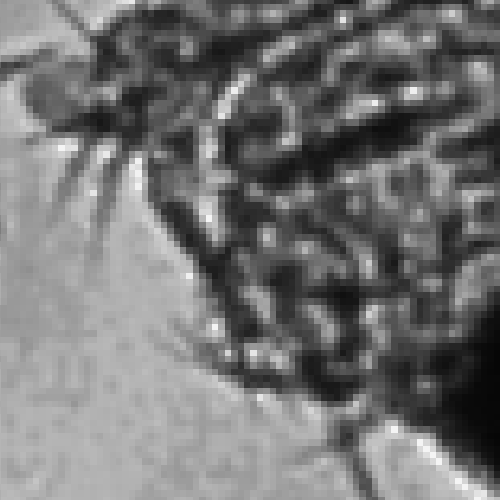}}}{\stackinset{l}{0pt}{t}{0pt}{\frame{\includegraphics[width=22pt]{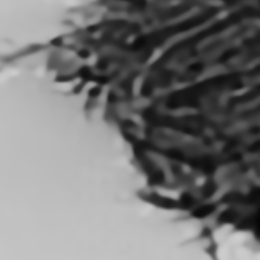}}}
    {\frame{\includegraphics[width=76pt, height=60pt]{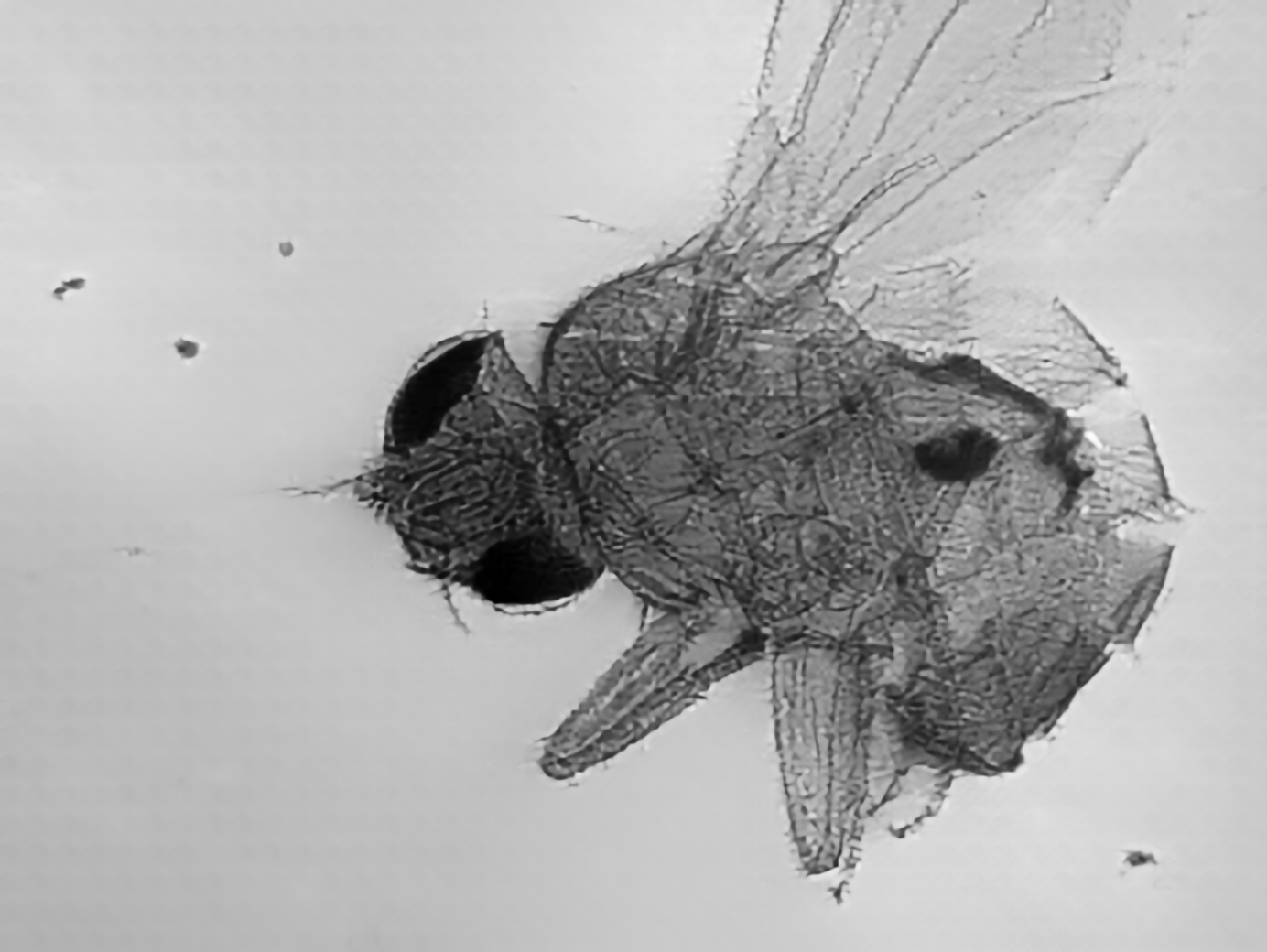}}}}
    }\\
    \subfloat[\texttt{Small intestine}, $40\times$ objective]{
    \frame{\includegraphics[width=76pt, height=60pt]{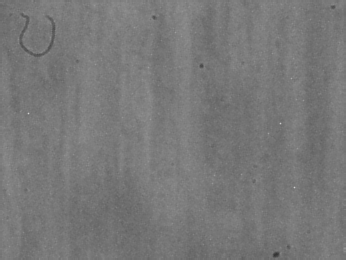}}
    \frame{\includegraphics[width=76pt, height=60pt]{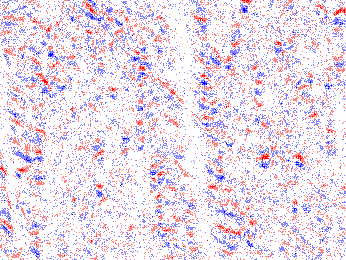}}
    \frame{\includegraphics[width=76pt, height=60pt]{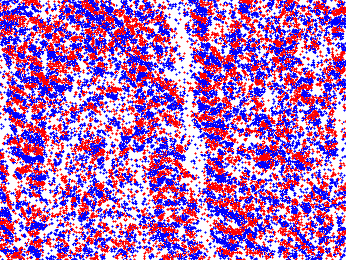}}
    \frame{\includegraphics[width=76pt, height=60pt]{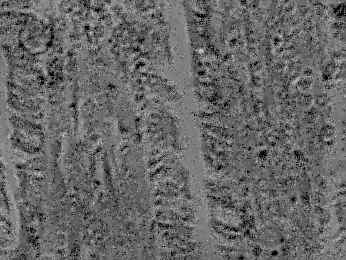}}
    }
    \end{tabular}
    \caption{(a) Neuromorphic microscopy system. (b)--(d) From left to right: raw image, raw LR events, SR events, and reconstructed image.}
    \label{fig:ms_sr}
\end{figure*}
\subsubsection{Neuromorphic Microscopy}
\textit{Why don't we use a larger magnification objective for imaging yet resort to SR algorithms?} Since there is a trade-off between the field of view and the desired magnification. Traditional frame-based microscopy is subject to limited temporal resolution and a low dynamic range, resulting in inferior observation of live specimens and dynamic processes that might exhibit low-contrast and blurring due to continuous motion. The two issues can be solved by the nature of neuromorphic imaging. As such, seeing and encoding the microscopic world through streaming events of neuromorphic microscopy is a promising alternative. As shown in~\cref{fig:ms_sr}, our imaging system exploits an upright widefield microscope (Nikon, Eclipse Ni-U) and a neuromorphic camera (iniVation, DAVIS346, $346 \times 260$ pixels, APS equipped) to capture frame- and event-based micrographs under a \SI{100}{\W} Halogen lamp illumination:
\begin{enumerate}
    \item \texttt{Honeybee hindleg}. The scopa, an essential apparatus on the tibia for carrying pollen, is our ROI. In the raw image, it exhibits quite low-contrast against the background that has high-contrast against the tibia. Some blurring caused by the moving hindleg is also observed. Due to weak motion and limited spatial resolution, the captured events are too insufficient in quantity to convey discernible components. We use the proposed method to obtain richer subpixel events, which are then integrated into the raw one to reconstruct a blur-free, high-contrast micrograph with much finer scopa texture details.

    \item \texttt{Drosophila}. The mouth and antennas making up olfactory organs is our ROI, shown in a zoom-in view. Few details are discernible in the raw image (top), and they cannot be recovered by frame $4\times$ SR (down). We perform $4 \times$ neuromorphic SR on the collected event stream, which is then transformed into an image with an equivalent resolution. Compared with the raw, the reconstructed image (top) clearly reveals the finer drosophila mouth and antennas faithful to the ground truth (down) taken with a $4 \times$ objective.

    \item \texttt{Small intestine}. We observe a small intestine tissue with a $40\times$ objective. Increasing the magnification leads to a reduced field of view and accordingly decreased luminous flux. Therefore, the imaging of a frame has to require a long exposure time of at least \SI{1}{\s}, and any slight movement during this period can bring severe blur, as the raw image shows. Being free of blur, neuromorphic imaging generates events within ultra-fast \SI{10}{\us} that contain a rough visual structure. Fusing the inferior image with the SR events of richer features can restore a complete, observable small intestine tissue.
\end{enumerate}
As collecting a huge volume of micrographs and their events is laborious and even unfeasible, the self-supervised mechanism is particularly suitable for handling such rare instances, whose parameters and conditions are quite different from natural images or synthetic samples on which current learning-based methods are trained. Supervised solutions, trained on a fixed configuration, are unlikely to perform well on the degradation or acquisition settings they have not ever seen, typically yielding unsatisfactory results. As such, self-supervised learning is a superior solution in these particular scenarios.

\subsection{Alternative to High-Resolution Events from Cameras}
While modern cameras often boast a high spatial resolution, the killing advantages of using HR neuromorphic cameras for sensing still remain a subject of debate. This is due to the significant challenges they pose --- the substantially higher demand for bandwidth, computing resources as well as the cost of hardware redesign. Besides, a previous study further backed such a counter-intuitive argument that, in scenes of low illumination and fast speeds, LR cameras significantly outperform HR ones that have a higher event rate, where the latter often brings a higher level of systematic noise, skipping events, timestamp perturbations, and then degraded results. Nevertheless, events from HR cameras still enjoy superior task performance in most scenarios, since more events always come with more information~\cite{gehrig2022arxiv}. Fortunately, neuromorphic SR offers a potential solution where high-quality events from a LR camera can be in an equivalent HR state to present richer scene features in a low noise level.

\subsubsection{Simulation Settings}
We investigate whether SR events generated by our method can serve as a faithful alternative to HR events from a HR camera, as evaluated on downstream tasks. The simulator~\cite{hu2021cvprw} synthesizes multi-scale event streams with different spatial resolutions $H\times W$, where $H = W \in \{128, 346, 640\}$ for the same field of view across all resolutions. The cutoff frequency $f_{c}$ (in~\si{\Hz}), which controls the rate at which each pixel monitors brightness input, is set to $200$ and $50$ to simulate realistic daytime and nighttime conditions, respectively. The contrast threshold $c$ is set to $0.2$. Then, our prototype super-resolves the LR events to reach an equivalent resolution. We thus have three sets of samples for comparisons --- LR ($128\times 128$ pixels), HR ($346\times 346$, $640\times 640$ pixels), and SR ($346\times 346$, $640\times 640$ pixels) events. 

\begin{figure}[t]
    \hspace{-10pt}
    \subfloat[LR ($128\times 128$)]{
    \begin{tabular}[b]{c}
        \stackinset{l}{0pt}{t}{0pt}{\frame{\includegraphics[width=25pt,height=20pt]{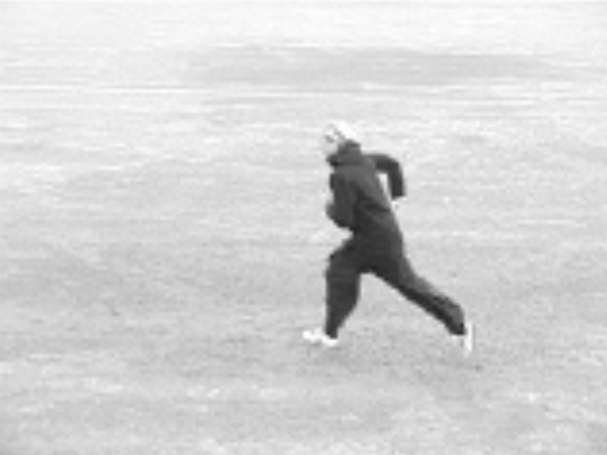}}}{\frame{\includegraphics[width=74pt, height=60pt]{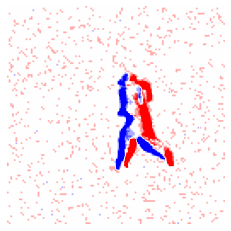}}}\\
        \stackinset{l}{0pt}{t}{0pt}{\frame{\includegraphics[width=25pt,height=20pt]{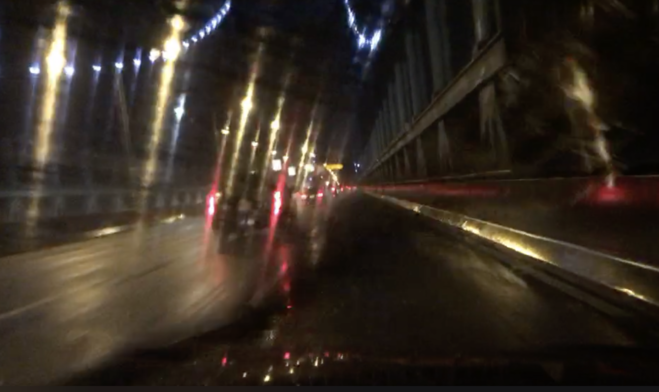}}}{\frame{\includegraphics[width=74pt, height=60pt]{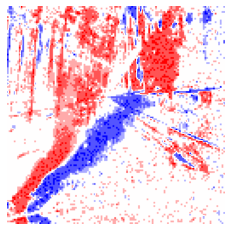}}}\\
    \end{tabular}}\hspace{-10pt}
    \subfloat[HR ($640\times 640$)]{
    \begin{tabular}[b]{c}
        \frame{\includegraphics[width=74pt, height=60pt]{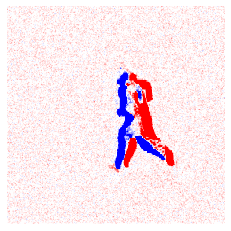}}\\
        \frame{\includegraphics[width=74pt, height=60pt]{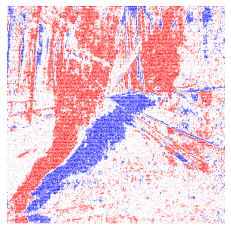}}\\
    \end{tabular}}\hspace{-10pt}
    \subfloat[SR ($640\times 640$)]{
    \begin{tabular}[b]{c}
        \frame{\includegraphics[width=74pt, height=60pt]{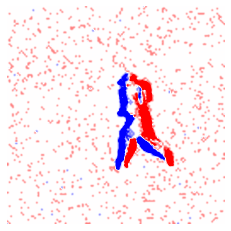}}\\
        \frame{\includegraphics[width=74pt, height=60pt]{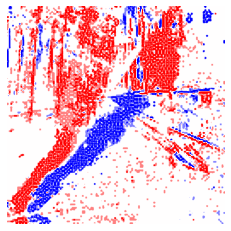}}\\
    \end{tabular}}
    \caption{Simulated LR and HR events from \texttt{daytime running}~\cite{schuldt2004icpr} and \texttt{nighttime driving}~\cite{yu2020cvpr}, along with the SR estimates by our approach.}
    \label{fig:lrhrsr_vis}
\end{figure}
\subsubsection{Visual Comparisons}
With the above setup, we simulate daytime and nighttime scenes at LR and HR scales, as shown in~\cref{fig:lrhrsr_vis}. In each case, the HR sample has a storm of noise in the background, while our SR result boasts an equivalent resolution and clarity yet maintains as a low noise level as the LR events, enjoying the features from the two sources.

\begin{table}[t]
\caption{Quantitative evaluations on LR, HR, and SR events.}
\label{table:lrhrsr}
\centering
\begin{tabular}{llcccc}
    \toprule
    \textbf{Data} & \textbf{Setting} & \multicolumn{4}{c}{\textbf{Evaluation}} \\ \midrule
    & & \multicolumn{2}{c}{\makecell{Image Reconstruction \\ (LPIPS~$\downarrow$)}} & \multicolumn{2}{c}{\makecell{Optical Flow Est. \\ (RNEPE~$\downarrow$)}}\\\cmidrule(lr){3-6}
     & & $f_c=200$ & $f_c=50$ & $f_c=200$ & $f_c=50$\\ \cmidrule(lr){3-6}
    \multirow{4}{*}{\texttt{running}} & $128$ LR & $0.38$ & $0.54$ & $1.98$ & $4.02$\\
    &$346$ HR & $0.33$ & $0.50$ & $1.43$ & $4.25$\\
    &\cb$346$ SR &\cb$0.30$ &\cb$0.44$ &\cb$1.51$ &\cb$4.06$\\
    &$640$ HR & $\mathbf{0.27}$ & $0.53$ & $1.66$ & $5.91$\\
    &\cb$640$ SR &\cb$0.28$ &\cb$\mathbf{0.40}$ &\cb$\mathbf{1.39}$ &\cb$\mathbf{3.83}$\\ \midrule
    \multirow{4}{*}{\texttt{toy}} & $128$ LR & $0.62$ & $0.78$ & $2.13$ & $5.63$\\
    &$346$ HR & $0.54$ & $0.61$ & $1.82$ & $5.92$\\\
    &\cb$346$ SR &\cb$0.51$ &\cb$0.62$ &\cb$1.65$ &\cb$5.43$\\
    &$640$ HR & $\mathbf{0.45}$ & $0.65$ & $1.80$ & $5.78$\\\
    &\cb$640$ SR &\cb$0.45$ &\cb$\mathbf{0.57}$ &\cb$\mathbf{1.52}$ &\cb$\mathbf{5.37}$\\ 
    \bottomrule
\end{tabular}
\end{table}
\subsubsection{Downstream Tasks} 
\Cref{table:lrhrsr} measures the event quality based on image reconstruction and optical flow estimation. For image reconstruction in daytime scenes, we observe a minimum at HR events, while they are outperformed by lower-resolution events in nighttime. Then, we exploit the resolution-independent normalized end-point-error (RNEPE)~\cite{gehrig2022arxiv} to compare predicted and ground truth flow. Consistent with the prior findings, HR events exhibit a limited advantage in optical flow estimation and perform poorly in nighttime scenarios. In the two tasks, our SR estimates from $128\times128$ LR events, elevate the performance to achieve comparable and even the best scores. Our approach, which can output high-quality SR events that boast the reduced-noise and rich-feature strengths from both sources, has the ability to be an effective simulation alternative to current HR cameras. 

\begin{figure}[t]
    \centering
    \includegraphics[width=0.48\textwidth]{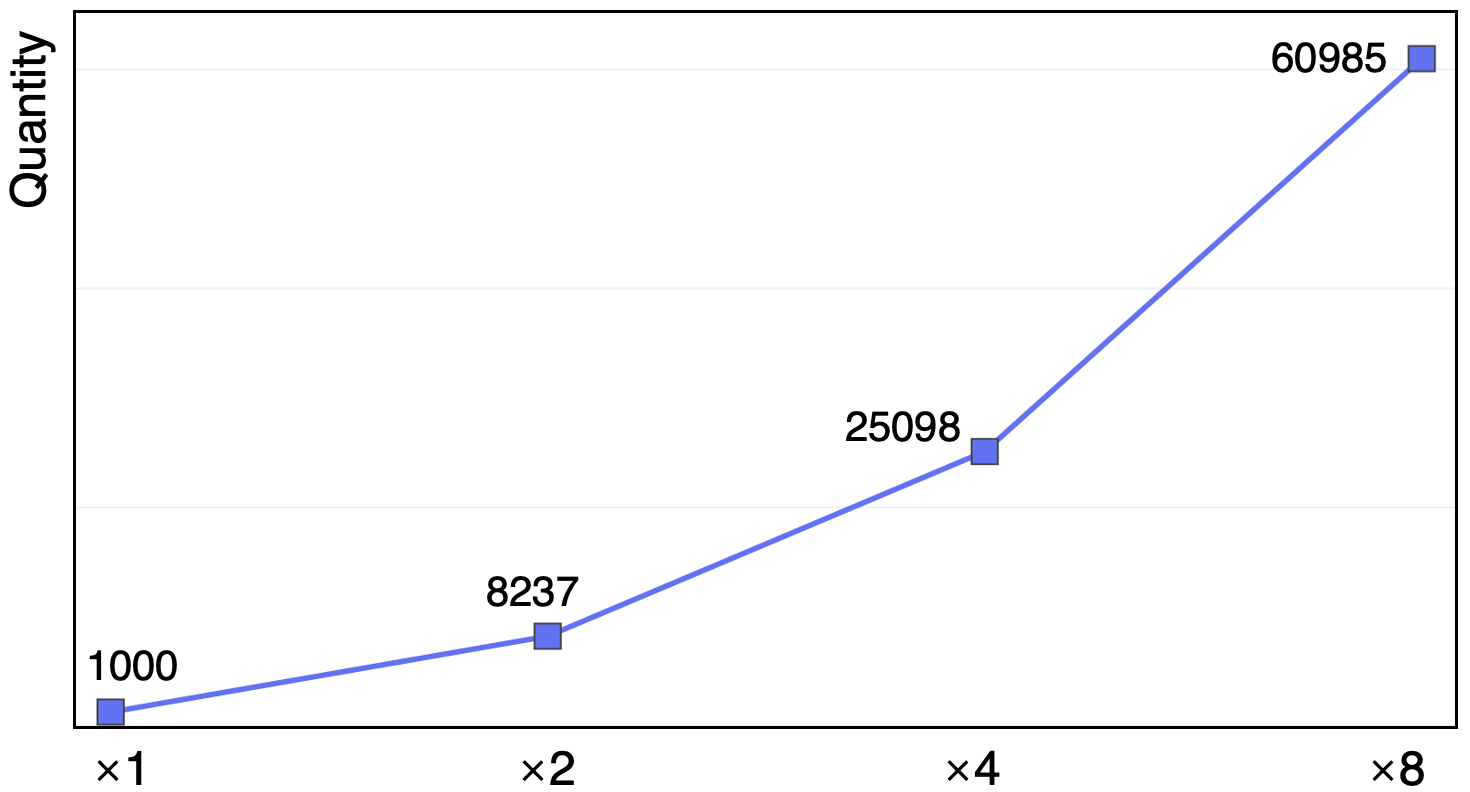}
    \caption{Event quantity dramatically increases at large-scale SR.}
    \label{fig:limitation}
\end{figure}
\subsection{Limitation Discussions}
Despite the convincing results achieved by our method, its known limitations should also be noted. As shown in~\cref{fig:limitation}, we compare the event quantity from different SR scales. The raw sample has only $1000$ events yet explodes to $60000$ at $8\times$ SR. It can be inferred that when on a large base number, high-scale SR might result in a huge data volume that demands much more processing delays. Current HR devices normally use complicated hardware-integrated filters to optimize the event rate. The focus of our work is on the exploration of a possibility that realizes neuromorphic SR in a self-supervised way. Integrating advanced techniques to filter out less informative events will be a direction for future research. Another research question that has not been thoroughly discussed in our work is the necessity of neuromorphic SR in various use cases. For example, \cref{fig:sr_reasoning}~(a) finds fewer gains obtained for the data with a higher noise level; \cref{fig:sr_reasoning}~(b) reveals that the larger-scale $4\times$ SR does not have a significant grow as $2\times$ SR; \Cref{table:lrhrsr} also shows slight improvements achieved by our SR events in some cases. Despite that neuromorphic SR is beneficial when current HR cameras are far from expectation, its necessity and usage scenarios still deserves more investigations for trade-offs between computing resources and desired performance.

\section{Conclusion}
Despite featuring microsecond temporal precision, neuromorphic imaging falls short in spatial resolution and presents a compromised level of visual clarity. This work proposes the first self-supervised learning prototype for neuromorphic SR, by which events are expanded and enriched along both spatial and temporal dimensions. Extensively assessed on downstream applications, this simple yet effective approach can acquire quite competitive results against the state-of-the-arts, significantly elevating flexibility without sacrificing accuracy. Given the limitations of current HR neuromorphic cameras and the ongoing debate surrounding their use in imaging, our solution becomes a cost-efficient and practical option.

\bibliographystyle{IEEEtran}
\bibliography{reference}
\newpage
\section{Biography Section}
\vskip -25pt plus -1fil
\begin{IEEEbiography}[{\includegraphics[width=1in,height=1.25in,clip,keepaspectratio]{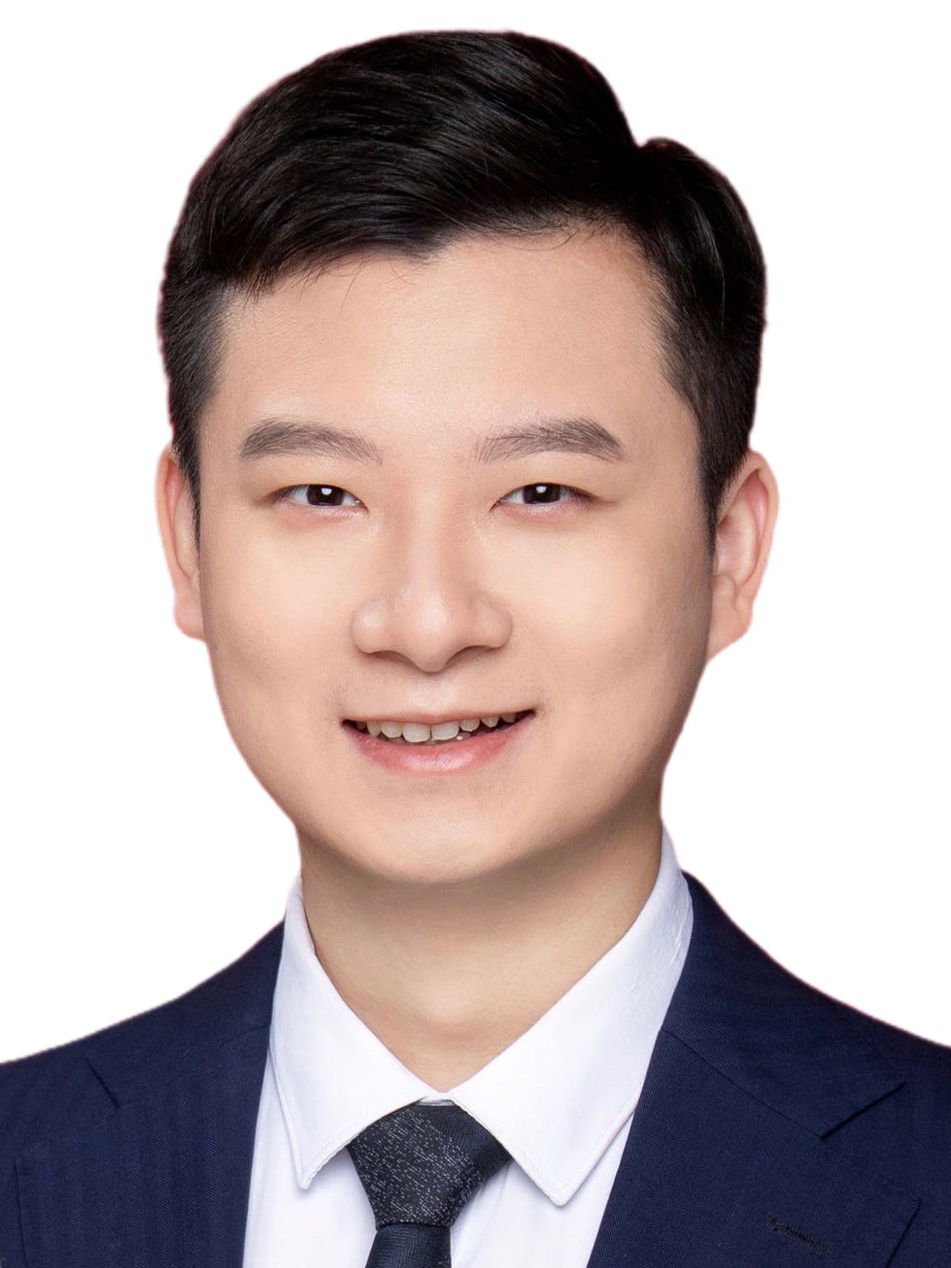}}]{PEI ZHANG} (Member, IEEE) received the B.Eng. degree from Beijing University of Posts and Telecommunications, in 2019, the B.S. degree from Queen Mary University of London, in 2019, and the M.S. degree from University College London, in 2020. He is currently pursuing the Ph.D. degree with the Department of Electrical and Electronic Engineering, The University of Hong Kong. His research interests include computational imaging, neuromorphic imaging and event-based vision.
\end{IEEEbiography}
\vskip -25pt plus -1fil
\begin{IEEEbiography}[{\includegraphics[width=1in,height=1.25in,clip,keepaspectratio]{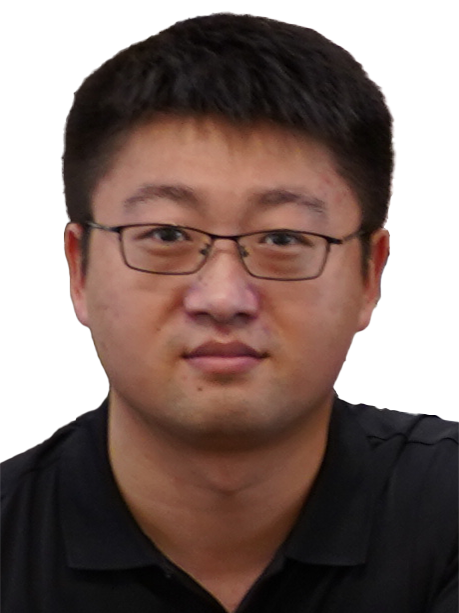}}]{SHUO ZHU} (Member, IEEE) received his B.S. degree from the Changchun University of Science and Technology in 2016, M.S. degree from the University of Shanghai for Science and Technology in 2019, and Ph.D. degree in optical engineering from the Nanjing University of Science and Technology in 2023. He is now a postdoctoral fellow at the University of Hong Kong. His research interest is computational neuromorphic imaging and its optical applications.
\end{IEEEbiography}
\vskip -25pt plus -1fil
\begin{IEEEbiography}[{\includegraphics[width=1in,height=1.25in,clip,keepaspectratio]{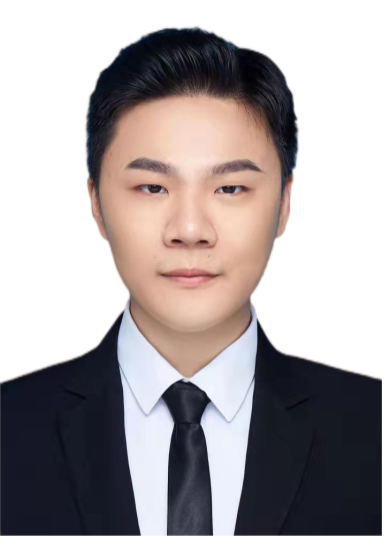}}]{CHUTIAN WANG} received the B.S. degree in Huang Kun Elite Class from the University of Science \& Technology Beijing in 2020, and the M.S. degree in the major of Optics and Photonics in Imperial College London in 2021. He was a research assistant in Zhejiang University until 2022. He is currently working towards his PhD degree with the Department of Electrical and Electronic Engineering, University of Hong Kong. His research interests include computational imaging and neuromorphic imaging.
\end{IEEEbiography}
\vskip -25pt plus -1fil
\begin{IEEEbiography}[{\includegraphics[width=1in,height=1.25in,clip,keepaspectratio]{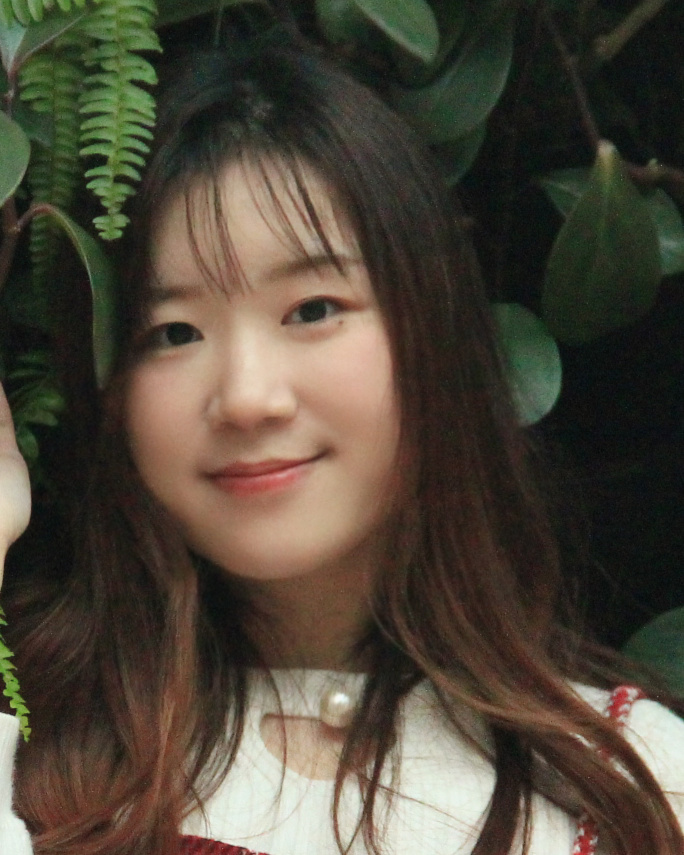}}]{YAPING ZHAO} (Student Member, IEEE) is currently a Ph.D. candidate in the Department of Electrical and Electronic Engineering at the University of Hong Kong (HKU), supervised by Prof. Edmund Y. Lam. Prior to joining HKU, she completed her Bachelor's degree at Beihang University under the supervision of Prof. Jichang Zhao in 2018, and her Master's degree at Tsinghua University under the supervision of Prof. Lu Fang in 2021, and visited Westlake University under the supervision of Prof. Xin Yuan in 2021.
\end{IEEEbiography}
\vskip -25pt plus -1fil
\begin{IEEEbiography}[{\includegraphics[width=1in,height=1.25in,clip,keepaspectratio]{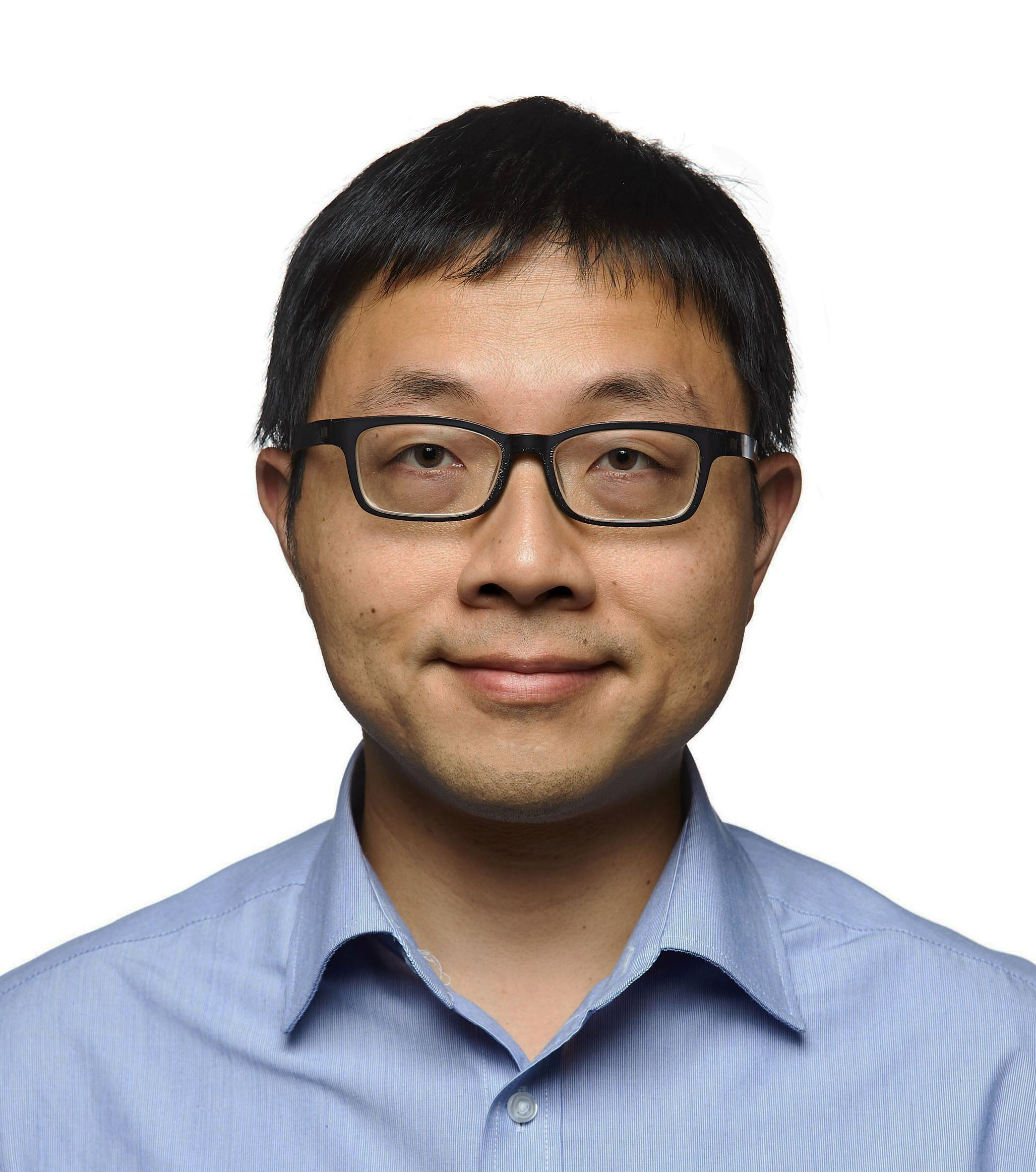}}]{EDMUND Y. LAM} (Fellow, IEEE) received the B.S., M.S., and Ph.D. degrees in electrical engineering from Stanford University. He was a Visiting Associate Professor with the Department of Electrical Engineering and Computer Science, Massachusetts Institute of Technology. He is currently a Professor of electrical and electronic engineering at The University of Hong Kong. He also serves as the Computer Engineering Program Director and a Research Program Coordinator with the AI Chip Center for Emerging Smart Systems. His research interest includes computational imaging algorithms, systems, and applications. He is a fellow of Optica, SPIE, IS\&T, and HKIE, and a Founding Member of the Hong Kong Young Academy of Sciences.
\end{IEEEbiography}

\end{document}